\documentstyle[12pt]{article}
\newcommand{\lwig}{\mbox{\,\raisebox{.3ex}
{$<$}$\!\!\!\!\!$\raisebox{-.9ex}{$\sim$}\,}}
\newcommand{\gwig}{\mbox{\,\raisebox{.3ex}
{$>$}$\!\!\!\!\!$\raisebox{-.9ex}{$\sim$}}\,}
\begin{document}
\begin{titlepage}
\begin{flushright} UCLA 94/TEP/31
\end{flushright}
\renewcommand{\thefootnote}{\fnsymbol{footnote}}
\begin{center}
{\bf \LARGE  Unifying Logarithmic And Factorial Behavior \\
in High Energy Scattering } \\
\vspace{0.25cm}
\baselineskip=18pt
{{\bf J.M. Cornwall\footnote[1]{cornwall@physics.ucla.edu}
and D.A. Morris}\footnote[1]{dmorris@physics.ucla.edu}  \\
University of California, Los Angeles\\
405 Hilgard Ave., Los Angeles, CA 90024,  U.S.A. \\
}
\date{}
\end{center}
\renewcommand{\thefootnote}{\arbic{footnote}}
\setcounter{footnote}{0}
\vspace{.25cm}
\thispagestyle{empty}
\begin{abstract}\normalsize
\noindent
The elegant instanton calculus of Lipatov
and others used to find factorially-divergent behavior
($g^N N!$) for $ Ng \gg 1 $ in $g \phi^4$ perturbation
theory is strictly only applicable when all external
momenta vanish; a description of high-energy
$2\rightarrow N$ scattering with $N$ massive particles
is beyond the scope of such techniques. On the other hand,
a standard multiperipheral treatment of scattering
with its emphasis on leading logarithms gives a reasonable
picture of high-energy behavior but does not result in
factorial divergences. Using a straightforward graphical
analysis we present a unified picture of both these phenomena
as they occur in the two-particle total cross section
of $g \phi^4$ theory.
We do not attempt to tame the unitarity violations associated
with either multiperipheralism or the Lipatov technique.
\vspace{.5cm}
\end{abstract}
\baselineskip=18pt
\begin{flushleft} UCLA 94/TEP/31\\
August 1994
\end{flushleft}
\end{titlepage}

\section{Introduction}
\baselineskip=18pt
\hspace*{\parindent}
There is a long-standing problem in high-energy scattering
which could have been addressed over a decade ago but was not,
apparently for a lack of applications. The problem resurfaced
a few years ago in a mutated form  when Ringwald\cite{ring}
and Espinosa\cite{esp} pointed out that a strict application
of the standard dilute-instanton-gas approximation (DIGA) in the
electroweak standard model leads to large $B+L-$violating
cross sections for $2\rightarrow N$ scattering when
$N \gwig O( 1 /\alpha_W )$ (that is, at energies
$\gwig O( M_W/\alpha_W) \simeq$ sphaleron mass\cite{mant}).
Shortly thereafter one us pointed out\cite{corn} that the same
DIGA, in the remarkable application of Lipatov\cite{lip}
and others\cite{bre} to finding the terms of large-order perturbation
theory\cite{gold}, gives the wrong high-energy behavior to purely
perturbative processes essentially because the DIGA cuts off
instanton size scales in the infrared by masses and
not by characteristic energies.

The Lipatov-DIGA provides a convenient means of obtaining
factorially divergent matrix elements
$T_{2 \rightarrow N}$ (for $N \gg 2$) which ultimately
need to be tamed by unitarity or, where applicable,
Borel summation\cite{zak}. In this paper we will not be concerned with
how these divergences are brought under control, which is an
interesting and important issue by itself,
but rather we will concentrate on the
their energy dependence --- an aspect which the Lipatov-DIGA
is not well suited to address.

The shortcomings of the Liptov-DIGA become evident
if one considers the dependence of
$T_{2 \rightarrow N}$ on the center
of mass scattering energy $E.$ When $E$ and the average energy
per outgoing particle, $E/N,$ are large compared to all masses
involved, one anticipates that $T_{2\rightarrow N}$ should scale
like $E^{2-N}$ (aside from logarithms and $N-$dependent factors).
However, in the Lipatov-DIGA $T_{2\rightarrow N}$ scales like
$M^{2-N}$ for some fixed mass $M,$ even if $E/N \gg M.$ The
suspicious energy-independence of the basic Lipatov-DIGA
calls into question the r\^{o}le of energetic
external particles in factorially divergent scattering amplitudes
and leads one to ask how such amplitudes can co-exist with
energy-dependent
amplitudes which do not diverge factorially. In this paper we
present a unified
analysis of this problem based on conventional Feynman diagram
techniques which demonstrate the emergence of factorially
divergent amplitudes and their interplay with familiar energy-dependent
amplitudes.

For definiteness, we will study the two-particle total
cross section in a theory characterized by the interaction Lagrangian
${\cal L}_I = -g \phi^4 / 4!.$ In particular, we will concentrate
on the imaginary part of the forward scattering amplitude
$Im~T_{2\rightarrow2}(t=0,s \gg m^2)$ which is related to the two-particle
total cross section through unitarity. In addition to introducing
a class of amplitudes whose sum exhibits factorially divergent
behavior we will also consider the consequences of amplitudes with
high-energy infrared (IR) logarithms (few graphs,
but with many partial waves) which profoundly alter
the overall energy dependence. These IR logarithms, associated with the
occurrence of many propagators with small momentum transfers
arise from graphical analyses much older than the Lipatov approach:
multiperipheralism\cite{multi}.

The study of multiperipheral graphs like
the uncrossed ladder graph of Fig.~\ref{fi:ladder}
is based on considering large orders of perturbation theory with an
emphasis on summing leading infrared logarithms which
eventually appear as sums of powers of $g\ln(s/m^2).$
Contributions of $N^{\rm th}$ order ladder graphs to
$Im~T_{2\rightarrow2}(t=0,s \gg m^2)$ are not accompanied
by a factor of $N!$ but rather by a factor of $1/N!$ and
these sum to give\cite{sawyer,arb}
\begin{equation}
\label{eq:arb}
Im~T_{2\rightarrow2}( s \gg m^2 , t = 0 ) \sim
\displaystyle{g^2 \over \ln^{3/2}\left( s/m^2 \right)  }
\left[
\left(           \frac{s}{m^2}          \right)^{
g/(16 \pi^2)}
+ ( g \rightarrow -g )  \right] \,\,.
\end{equation}
The term with the sign of $g$ reversed ensures that only
even powers of $g$ appear in the final answer. In contrast to
ladder graphs, the class of graphs which we will introduce,
whose sum gives $N!$ behavior, is free from logarithms.

The characteristic features of $g \phi^4$ diagrams without logarithms
in their imaginary part are the absence of multiple lines
between vertices and the absence of lines beginning and ending
on the same vertex (tadpoles). In graph-theoretic language
such diagrams are known as ``simple graphs''.
In section~2 we introduce a subset of simple
graphs which we call $K-$graphs whose properties are easily
understood since they are straightforward generalizations of
multiperipheral graphs; they have neither infrared nor ultraviolet
logarithms in their imaginary part. A crucial part of our analysis is a
factorization theorem, stated in Section~3 and proved in
appendix~D, which allows us to isolate the contribution of
simple graphs from diagrams where they mix with ladder graphs.
Though we illustrate the factorization theorem using $K-$graphs,
the theorem holds for all simple graphs.

Using factorization to combine simple graphs with ladder graphs
gives an extremely concise result: in the analysis of
ladder graphs one effectively makes the substitution
\begin{equation}
\label{eq:gsub1}
g \rightarrow g \left(
1 + \sum_{{\rm even}~N} \left( \frac{a g}{16 \pi^2 }\right)^N N!
\right)^{1/2}
\end{equation}
where the sum on the right-hand side is recognized as being
factorially divergent. Specifically, Eq.~\ref{eq:arb} is
modified to
\begin{equation}
\label{eq:second}
Im~T_{2\rightarrow2}( s \gg m^2 , t = 0 )
\sim
\displaystyle{
g^2 \beta(g) \over
\ln^{3/2} \left( \displaystyle{s\over m^2} \right)  }
\left[ \left( \frac{s}{m^2} \right)^{\alpha(g)}
 + \,\,\, ( g \rightarrow -g ) \,\,\right] \, \, ,
\end{equation}
where
\begin{equation}
\label{eq:dotdef}
\beta (g) = \left( \displaystyle{16 \pi^2 \over g } \alpha(g) \right)^2
\doteq
1 + \sum_{{\rm even}~N} \left( \frac{a g}{16 \pi^2 }\right)^N N!
\,\, .
\end{equation}
The difference between considering just $K-$graphs
or all simple graphs amounts to different estimates for
the constant $a.$ Including only $K-$graphs we find the
bound $a_K \ge 1/2$ and we estimate $a \simeq 0.94$
if all simple graphs are included; the
corresponding Lipatov value is $a=1.$
The difference between our estimates and the Lipatov value
likely reflects the fact that our graphical
sum of contributions to $Im~T_{2 \rightarrow 2}$
is not exhaustive.

In Eq.~\ref{eq:dotdef} we have introduced a modified equality sign
$\doteq$ which we use throughout the paper. Two $N-$dependent
functions $A_N,$ $B_N$ (or sums of such functions) obey
$A_N \doteq B_N$ if, for large $N,$ $A_N$ and $B_N$ are
of the form $c_1 ( N! )^{ c_2 } c_3^N N^{c_4} ( 1 + O(1/N) )$
with $c_2$ and $c_3$ being the same constants
for both $A_N$ and $B_N.$  That is, overall constant factors,
fixed powers of $N,$ and non-leading terms are ignored,
and may differ from $A_N$ to $B_N.$

For non-forward scattering $(t\ne 0)$
the terms in the sum of Eq.~\ref{eq:dotdef} have
extra factors $F_N(t)$, with $F_N(0)=1,$ which are calculable in principle
with our technique, but we will not consider the case $t\ne 0$ here.
Note, by the way, that we only treat graphs corresponding to even powers of
$g$ whereas the Lipatov analysis treats both even and odd powers.
Consequently, $(-1)^N$ factors which would normally appear
in a sum like Eq.~\ref{eq:dotdef} have no effect.

	One might think that the result of
Eqs.~\ref{eq:second}-\ref{eq:dotdef} is obvious and,
in a rather hand-waving way, it is.
The problem, however, is to find a way to derive it
and we know of no direct application of the Lipatov technique
for doing this. Our approach simply involves summing graphs, sometimes
using the Bethe-Salpeter equation. Though there are many
regularities in our analysis which suggest an underlying
semi-classical behavior, we cannot yet assemble these clues
into a Lipatov-like semi-classical derivation of our results.

The difficulty in obtaining the true high-energy behavior from
Lipatov theory can be traced to the fact that the Lipatov-DIGA
analysis is, in principle, only applicable to graphs with zero
four-momentum on every external leg\cite{corn2}. This is a consequence
of the vanishing of all the matrix elements of the semi-classical
energy momentum tensor when evaluated on the Lipatov instanton-like
solutions of wrong-sign $g\phi^4$ theory.
One might hope that summing quantum corrections
to the basic DIGA result will make it possible to derive high-energy
results, but so far work\cite{mattis}
on the analogous problem of high-energy B+L violation has not yet
resolved this problem. What is needed for the semi-classical analysis to
succeed at high energy is to find
a means of communicating the energy of the
external on-shell particles to the instantons,
a feat which is most easily accomplished in Minkowski space.
Otherwise, the instantons and analogous objects in the Lipatov
analysis can be of any size and, when all sizes are integrated over, the
$M^{2-N}$ behavior for $T_{2\rightarrow N}$
emerges, as mentioned earlier. There has been speculation
as to how information on external energies can be transferred to
the instantons by solving inhomogeneous instanton equations with
external-particle energy-dependent sources\cite{corn2}
but exact solutions of the relevant equations do not
exist with the necessary generality.

	In the graphical approach, all graphs begin with the
appropriate on-shell high-energy external lines attached. Certain
graphs, or parts of graphs, can be factored out or otherwise
identified in the overall process, which have
the property that they are independent of the external
momenta, at least for forward scattering; such graphs
look  very much like graphs with all four-momentum
vanishing even though they are Minkowski-space
high-energy graphs, or parts thereof. It is these graphs
which contribute to $\alpha,$ $\beta$ and the
$N!$ behavior of Eq.~\ref{eq:second}. They are
also the graphs for which we would expect some sort of
semiclassical analysis to hold,
but as mentioned above we do not know how to do this at the level
of $Im~T_{2\rightarrow2}$ itself.

	An interesting problem for the future would be to find
a semiclassical analysis for $Im~T_{2 \rightarrow 2}$ at
high energy. A non-trivial extension of the present paper is
to generalize the results on $Im~T_{2 \rightarrow 2}$ to multiloop
graphs for $T_{2 \rightarrow N}.$ The most challenging problem is
to extend the results to gauge theories, where new complications
arise. In $g \phi^4$ theory the reduced graphs
contributing to $\alpha$ and $\beta$ of Eq.~\ref{eq:second} are
four-dimensional while in gauge theory they are two-dimensional\cite{cheng}.
Since there are infinitely many conformally invariant
two-dimensional theories, there is no standard Lipatov analysis for such
theories. Recently, there
has been speculation\cite{verlind} on the form of this
two-dimensional theory for non-abelian gauge theories and gravity
in four dimensions, but the problems of incorporating the
correct high-energy behavior (i.e., the leading logarithms) has
only begun to be addressed in this context. We will briefly discuss
these problems for the future in the concluding section.

The outline of the paper is as follows. In Sect.~2 we identify
a class of diagrams which generalizes the notion of multiperipheral
graphs and we demonstrate that their contribution to
$Im~T_{2 \rightarrow 2}$ gives rise to an energy-independent
$N!$ behavior analogous to that found by Lipatov-DIGA techniques.
We call the generalized graphs $K-$graphs since
we eventually use them to define a kernel for a Bethe-Salpeter equation.
The statistical picture which emerges in Sect.~2 brings to mind the work
of Bender and Wu\cite{ben76} and Parisi\cite{par77}. Our analysis goes
beyond previous studies in that we keep track of the manner in which
$K-$graphs exhibit factorial divergences so that we can later
merge them with energy-dependent amplitudes. In Sect.~3 we
form chains of $K-$graphs and two-line loops (which by themselves
are responsible for leading-log behavior) and sum them to obtain
the result of Eqs.~\ref{eq:second}-\ref{eq:dotdef}. We uncover
a factorization theorem which reduces the problem of summing
chains of $K-$graphs and two-line loops to the well-studied
problem of summing straight-ladder graphs.
Details of our calculations are given in Appendices~A-E.

\section{Graphical Analysis}

\subsection{Choice of Graphs}

\hspace*{\parindent}
Consider the uncrossed ladder graph of Fig.~\ref{fi:ladder}.
Since there is only one such graph in the $N^{\rm th}$ order
of perturbation theory its statistical weight can not contribute
to the $N!$ behavior of $Im~T_{2 \rightarrow 2}.$
Nevertheless, the imaginary part of this graph has the maximum
possible powers of $\ln( s/ m^2)$ (which is $N-2$ in $N^{\rm th}$ order).
Here $m$ is the mass of the particle propagating in the vertical lines of
Fig.~\ref{fi:ladder}; one can safely
ignore all other masses without encountering
additional IR divergences and we will do so whenever possible.
In particular, we take $p^2 = p^{'2} = 0.$ Since Fig.~\ref{fi:ladder} is
essentially the square of a tree graph, it will not contribute
UV logarithms to $Im~T_{2 \rightarrow 2}$ (though it contributes
a single UV logarithm to $Re~T_{2 \rightarrow 2}).$

One can generalize the notion of multiperipheral graphs by
drawing two opposing trees as in Fig.~\ref{fi:tree} (corresponding
to a $2 \rightarrow N$ contribution to
$Im~T_{2 \rightarrow 2} \sim \sum | T_{2 \rightarrow N} |^2$)
and then joining the lines on one tree to the lines on the other
in all possible ways. Among the graphs formed in this manner
are uncrossed ladder graphs like Fig.~\ref{fi:ladder}
(which have many two-line loops), graphs like Fig.~\ref{fi:k8}a
(which have no two-line loops), crossed ladder graphs
like Fig.~\ref{fi:4cross}a, and graphs such as those
of Fig.~\ref{fi:tangle} (which include graphs with a few two-line
loops). In this paper we will restrict our attention to two
types of graphs and a certain way of intertwining them.
The first type of graph is the familiar uncrossed
ladder of Fig.~\ref{fi:ladder} while the second type
has as its simplest example the graph of
Fig.~\ref{fi:tangle}a.
The defining features of graphs like
Fig.~\ref{fi:tangle}a are the absence of two-line loops
and two-particle irreducibility in $t$ (vertical) channel.

An algorithm for constructing $N^{\rm th}$ order graphs like
Fig.~\ref{fi:tangle}a begins by marking each of two vertical lines
with $N/2$ vertices ($N$ is even for all our graphs) as
in Fig.~\ref{fi:tree}. One then takes a closed loop of string and
attaches it to the vertices in such a way that in tracing the string
one alternately goes from one vertical line to the other without revisiting
any vertex. We will call graphs constructed in this manner $K-$graphs;
Fig.~\ref{fi:k8}a is an example of an eighth-order $K-$graph.
In Sect.~\ref{se:kgraphs} we show that there are $O(N!)$
$K-$graphs in the $N^{\rm th}$ order of perturbation theory and
that, due to their lack of two-line loops,
$K-$graphs contribute no logarithms to $Im~T_{2 \rightarrow 2}.$
Among the graphs we exclude by concentrating on $K-$graphs
are graphs like Fig.~\ref{fi:4cross}a (which has fewer logarithms than
the corresponding uncrossed ladder graph) and Fig.~\ref{fi:4cross}b
(which has UV logs). We also exclude graphs like Fig.~\ref{fi:tangle}c
which may affect the details of our quantitative results but which do
not change the overall picture.

There are, of course, many other graphs besides $K-$graphs which
are free of two-line loops and hence have no logarithms in their
imaginary part. In Appendix~E we invoke a far-reaching
combinatorial result of Bender and Canfield\cite{bencan}
which accounts for all such graphs. Despite its generality, however,
the Bender and Canfield result does not immediately lend itself to
an intuitive interpretation. For this reason we choose to illustrate
many of our arguments by using the less comprehensive class of
$K-$graphs. Though $K-$graphs represent the simplest generalization
of multiperipheral graphs, our analysis is equally applicable to all
logarithm-free graphs.

Before proceeding with our analysis of $K-$graphs, it is
worthwhile mentioning how the extensively-studied factorial
divergences\cite{corn,gold} in off-shell decay amplitudes
$T_{1 \rightarrow {\rm many}}$ fit into the present picture.
Consider the graph of Fig.~\ref{fi:cayley} in which two decay-like
trees emerge from the point where $p$ and $p'$ meet.
Summing over all such amplitudes and squaring yields
a contribution to the total cross section and consequently also
to $Im~T_{2 \rightarrow 2}.$  Suppose the tree-like decays
in Fig.~\ref{fi:cayley} terminate with the production of
$N_1$ and $N_2$ particles where $N_1 + N_2 = N.$
Assuming the equipartition of the center of mass energy $E$
among all $N$ particles, the sum over all Cayley tree graphs of
the form of Fig.~\ref{fi:cayley} results in a lower-bound
contribution to $T_{2 \rightarrow N}$
\begin{equation}
T_{2\rightarrow N}
\doteq \sum_{N_1 + N_2 = N} \, F_{N_1} \, F_{N_2} \,
\displaystyle{N! \over N_1! N_2! }  ,
\end{equation}
where\cite{corn}
\begin{equation}
F_{N_i}
\doteq
\left(
\displaystyle{ g \over 6 \alpha^2 }
\right)^{N_i /2}
N_i!
\left(
\displaystyle{ E \over N} \right)^{1-N_i}
\end{equation}
with $\alpha =2.92.$ For such graphs we estimate the contribution
\begin{equation}
Im~T_{2 \rightarrow 2} \doteq  |T_{2 \rightarrow N}|^2  \rho_N / N!
\end{equation}
where massless relativistic phase space is
\begin{equation}
\rho_N \doteq \left( \displaystyle{ E \over  4 \pi } \right)^{2 N - 4}
              \displaystyle{1\over (N!)^2}       .
\end{equation}
Putting everything together, the contribution
to $Im~T_{2 \rightarrow 2}$ from Cayley trees is
\begin{equation}
\label{eq:cayleyresult}
Im~T_{2 \rightarrow 2} \doteq N!
\left( \displaystyle{ a_C g \over 16 \pi^2 }
\right)^N  , \qquad \qquad
a_C
 = \displaystyle{e^2\over 6 \alpha^2} \simeq  0.14
\quad .
\end{equation}
Though Eq.~\ref{eq:cayleyresult} certainly implies factorially
divergent behavior, it turns out we will obtain much stronger
bounds on $a$ from $K-$graphs ($a_K \ge 1/2$) and ultimately,
in Appendix~E, when we consider all logarithm-free graphs, we
will bound $a \ge 2/3$ and estimate $a \simeq 0.94.$

\subsection{ Feynman-Parameter Representation of Graphs}

\hspace{\parindent}
\label{se:feynman}
This is a well-known subject\cite{tik}, and we will be brief.
Every graph we consider has just one UV logarithm in the real part,
so the momentum integral must be regulated.
This can be done with dimensional
regularization or by applying $\partial/\partial m^2 $ to the
graph before performing  the momentum integrals and then
integrating with respect to $m^2$ after taking the imaginary part.
Either way one finds that a $N^{\rm th}$ order graph (for $N$ even)
with symmetry factor $S$ contributes
\begin{equation}
\label{eq:imt}
Im~T_{2 \rightarrow 2}  = \displaystyle{g^2 \over 16 \pi }
       \left( \displaystyle{ g \over 16 \pi^2 } \right)^{N-2}
       \displaystyle{ 1 \over S }
\int^1_0 \displaystyle{ [dx] \over U^2 } \,\,\Theta\!\left(
\displaystyle{\phi \over U} - m^2 \right) \,\, ,
\end{equation}
where $[dx]$ is the usual measure associated with
an integral over $2 N- 2$ positive Feynman parameters $x_i,$
\begin{equation}
[dx]  =   dx_1 \cdots dx_{2N-2}\, \,
\delta\!\left( 1 - \sum x_i  \right) .
\end{equation}
$U$ is the sum of all products of $N-1$ Feynman
parameters such that cutting the corresponding lines
leaves a single connected tree graph. $\phi$ is
given by
\begin{equation}
\label{eq:phi}
\phi = s \phi_s + u \phi_u
\end{equation}
where $\phi_s$ $(\phi_u)$ is the sum of all products of $N$ Feynman
parameters such that cutting the corresponding lines
splits the original graph into two connected trees with
the square of the four-momentum entering each tree
being $s=(p+p')^2$,  $(u = (p-p')^2 ).$
In general, $\phi$ also contains terms proportional
to $t$ and $m^2$ but these may be ignored because
we are interested in forward scattering ($t=0$)
in the high-energy limit $(s \gg m^2)$.

	A crucial property of $U$ is that it cannot vanish unless
all of the parameters of a single loop vanish simultaneously.
Consider a $l-$line loop labelled by the parameters
$x_1,x_2, \dots, x_l.$  Inserting in  Eq.~\ref{eq:imt} the identity
\begin{equation}
\label{eq:scale0}
1 = \int_0^1 d\lambda \, \delta\!\left( \lambda - \sum_1^l x_i \right)
\end{equation}
and then performing the change of variables
\begin{equation}
\label{eq:scale}
x_i = \lambda \tilde{x}_i \, , \qquad \qquad i = 1, \dots l\,\,,
\end{equation}
one finds that $U$ vanishes linearly in $\lambda$ as
$\lambda \rightarrow 0.$  If the Feynman parameters of another
(possibly overlapping) loop are analogously scaled with
a parameter $\lambda',$ then $U \sim \lambda \lambda'$ as
$\lambda,\lambda' \rightarrow 0.$ Since $U^2 \sim \lambda^2$
when $\lambda$ is small, the $\lambda-$integral
in Eq.~\ref{eq:imt} is of the form $\int_0^1 \lambda^{l-3} d\lambda$
which gives logarithms only if $l=2.$ Consequently, requiring
an absence of two-line loops in a graph ensures an absence
of logarithms in the contribution to $Im~T_{2 \rightarrow 2}.$

\subsection{ $K-$graphs}
\label{se:kgraphs}

\hspace{\parindent}
    Here we establish several properties of $K-$graphs, culminating
in a lower bound on the contribution of all
$N^{\rm th}$ order $K-$graphs to $Im~T_{2 \rightarrow 2 }$
which is indeed $O(N!).$
Our bound is of the form that the naive
Lipatov analysis would give, although strictly
speaking Lipatov's methods are not applicable for
for high-energy Minkowski-space graphs.

    We begin by counting the number of $N^{\rm th}$ order $K-$graphs.
Consider again the construction of
two vertical lines or ``walls'' each of which
has $N/2$ vertices; in a $2\rightarrow 2$ multiperipheral
amplitude these walls would correspond to the spacelike
propagators: incoming momenta
would be attached to the top and bottom of one wall while the
outgoing momenta are similarly attached to the other wall. The
remaining internal lines of a $K-$graph may then be visualized as the
trajectory of a fictitious particle which ``bounces'' from wall
to wall such that each of the $N$ vertices is visited exactly once
before the trajectory closes --- this ansatz guarantees the
absence of two-line loops.

Finding the number of possible trajectories
in the above construction is a simple combinatoric exercise.
Consider a trajectory as it leaves, say,
the bottom vertex of the left wall:
there are $\frac{N}{2}$ possible vertices on the right wall
which the trajectory can intersect. After bouncing off the right
wall, the trajectory can return to any one of $\frac{N}{2} - 1$
unvisited vertices on the left wall. Similarly, after bouncing
off the left wall, there are $\frac{N}{2}-1$ unvisited vertices
on the right wall. Continuing in this manner, one finds
exactly $(\frac{N}{2})! ( \frac{N}{2} - 1)! $ possible
directed trajectories. Dividing this number by two
(to avoid double counting trajectories which are identical
except for the sense in which they are traversed)
we find that the number of $N^{\rm th}$ order $K-$graphs is
\begin{equation}
\label{eq:nkgraphs}
\displaystyle{1 \over 2}
 \left(
\displaystyle{N \over 2}
\right)!
\left( \displaystyle{N \over 2} - 1 \right)!
\doteq \displaystyle{ N! \over 2^{N} }
\end{equation}
where, as before, $\doteq$ means equality modulo constant factors
and fixed powers of $N.$

A consistent application of the Feynman rules
demands that a sum of contributions to $Im~T_{2\rightarrow2}$
of the form of Eq.~\ref{eq:imt} must include only topologically
inequivalent graphs. It turns out, however, that
not all $\frac{1}{2} (\frac{N}{2})! ( \frac{N}{2} - 1)! $
$K-$graphs are topologically distinct: the first sign of
redundancy appears at $N=10$ where the topologies
of $12$ graphs (out of 1440) appear twice. Fortunately,
this level of duplication is insignificant (in the $\doteq$ sense) and
so we are still
justified in identifying $N!/2^N$ in Eq.~\ref{eq:nkgraphs}
as the number of topologically inequivalent $K-$graphs.
Another minor point concerns the symmetry factor $S$
associated with each $K-$graph in Eq.~\ref{eq:imt}.
Since the fraction of $N^{\rm th}$ order
$K-$graphs with $S\ne1$ goes to zero
as $N$ grows asymptotically large\cite{bol82} we will
implicitly assume that all $K-$graphs have $S=1.$

We next direct our attention to the integral of Eq.~\ref{eq:imt}.
Using Eq.~\ref{eq:phi} and setting $u=-s,$ the $\Theta-$function
appropriate for a $K-$graph is
\begin{equation}
\label{eq:thetaapp}
\Theta\left(
  \displaystyle{ \phi_s - \phi_u \over U }
- \displaystyle{m^2 \over s} \right)
 \simeq \Theta \left( \phi_s  - \phi_u \right)
\end{equation}
where we have legitimately dropped the small
parameter $m^2/s \ll 1$ because there are no
logarithmic divergences to regulate.
Though the $\Theta-$function in Eq.~\ref{eq:imt} poses no
problem in principle, it can be eliminated if we consider
additional contributions to $Im~T_{2 \rightarrow 2}$
which are closely related to those of $K-$graphs.
To illustrate this point consider the $K-$graph of
Fig.~\ref{fi:k8}a and its associated $u-$channel exchange
graph of Fig.~\ref{fi:k8}b which is identical
except for switching the vertices to which the
ingoing and outgoing four-momentum vectors $p'$ are attached.
If the internal lines of
Figs.~\ref{fi:k8}a,b are labelled with the
same Feynman parameters then it
follows from the graph-based definition of $U$ and $\phi$ that
\begin{equation}
U^{\rm{Fig.~\ref{fi:k8}b}} = U^{\rm{Fig.~\ref{fi:k8}a}},  \qquad\qquad
\phi_s^{\rm{Fig.~\ref{fi:k8}b}} = \phi_u^{\rm{Fig.~\ref{fi:k8}a}},
\qquad\qquad
\phi_u^{\rm{Fig.~\ref{fi:k8}b}} = \phi_s^{\rm{Fig.~\ref{fi:k8}a}}
\, \, .
\end{equation}
In other words, the $\Theta-$functions associated
with Figs.~\ref{fi:k8}a,b sum to unity so that the corresponding
joint contribution to $Im~T_{2\rightarrow2}$ is
\begin{equation}
\label{eq:notheta}
Im~T_{2 \rightarrow 2}  = \displaystyle{g^2 \over 16 \pi }
       \left( \displaystyle{ g \over 16 \pi^2 } \right)^{N-2}
\int^1_0 \displaystyle{ [dx] \over U^2 }.
\end{equation}
It is clear that the $\Theta-$function for any $K-$graph can similarly
be eliminated by including the appropriate $u-$channel
exchange graph. Over-counting is not an issue
for the graphs of Figs.~\ref{fi:k8}a,b. When
Fig.~\ref{fi:k8}b is redrawn in Fig.~\ref{fi:k8}c
it is apparent that it is not one of the
original $\frac{1}{2} (\frac{N}{2})! (\frac{N}{2} - 1 )!$
$K-$graphs. There are, however, a small number
(in the $\doteq$ sense) of $K-$graphs whose exchange
graphs are themselves $K-$graphs. Though such graphs
pose no problem in principle, we
will neglect them in order to keep our discussion simple.

	If we implicitly agree to include $u-$channel exchange
graphs, all that remains is to evaluate
\begin{equation}
\label{eq:idef}
I_i = \int_0^1 \displaystyle{ [dx] \over  U^{2}_i }
\end{equation}
where $U_i$ is the $U$ function for $i^{\rm th}$ $K-$graph
and then to sum over all $N!/2^N$ $K-$graphs.
As far as we know it is impossible
to calculate $I_i$ analytically but one
can obtain useful lower bounds from the following simple considerations.
For each graph we define the normalized probability density $p_i$
for $U,$
\begin{equation}
\displaystyle{  p_i  }  =  \displaystyle{ \int^1_0 [dx]   \,
                 \delta( U - U_i(x) )
                 \over
                 \int^1_0 [dx]                } .
\end{equation}
It is not difficult to show that $ \int_0^1 [dx] = 1 / (2 N - 3 )! $
which allows us to write
\begin{eqnarray}
\label{eq:lowb}
I_i & = &
\displaystyle{ 1 \over ( 2 N - 3)! }
\int dU \displaystyle{ p_i \over U^2 } \nonumber \\
  & \equiv &
\displaystyle{ 1 \over ( 2 N - 3)! }
\left\langle \displaystyle{1 \over U^2 }
\right\rangle_i  \nonumber \\
& \ge &
\displaystyle{ 1 \over (2 N - 3)!}
\,\, \displaystyle{ 1 \over \langle U \rangle^2_i } \,\, ,
\end{eqnarray}
where $\langle 1/U^2 \rangle_i \ge 1/\langle U \rangle^2_i$
follows from the H\"{o}lder inequality\cite{hardy}
\begin{equation}
\int \, dU \, f g \, \ge \,
\left( \int \, dU \, f^k \right)^{1/k}
\left( \int \, dU \, g^{k'} \right)^{1/k'}
\end{equation}
with $f = U / \sqrt{p_i},\,\, g = (p_i)^{3/2},\,\, k = -2$
and $k' = k/(k-1) = 2/3.$

Our interest in the bound of Eq.~\ref{eq:lowb} arises from
the fact that $\langle U \rangle_i$ is easily calculated
for any given graph. Specifically, one has
\begin{equation}
\label{eq:averages}
\left\langle U \right\rangle_i
=
C_i \,\, ( 2 N - 3 )! \,
\int_0^1 [dx] \, x_1 x_2 \cdots x_{N-1}    \,\, ,
\end{equation}
where $C_i$ is the number of terms in $U_i.$ In graph-theoretic
language $C_i$ is known as the complexity or number of
spanning trees of a graph\cite{biggs}.
The integral in Eq.~\ref{eq:averages} is elementary; starting
from the Feynman identity
\begin{equation}
\prod^{2 N-2}_{i=1} \frac{1}{A_i} =
( 2 N - 3 )! \int [dx] \left( \sum A_i x_i \right)^{2 N - 2 }
\end{equation}
and differentiating once with respect to
$A_1, A_2, \cdots, A_{N-1}$ and then setting all the $A_i=1,$
one finds $\int_0^1 [dx] x_1 x_2 \cdots x_{N-1} = 1/(3N - 4)!.$
Summing over all $N^{\rm th}$ order $K-$graphs, we obtain
the bound
\begin{equation}
\sum_i  I_i \ge
\displaystyle{ (3 N - 4)!^2 \over (2N - 3)!^3  } \,\,\,
\sum_i \displaystyle{1 \over C_i^2}.
\end{equation}
We next introduce a normalized complexity density function
\begin{equation}
p_C \equiv \displaystyle{ 2^N \over N! } \sum_i
\delta( C - C_i )
\end{equation}
where again the sum is over all $N^{\rm th}$ order $K-$graphs.
In this language we have
\begin{equation}
\sum_i I_i \,\, \ge \,\,
\displaystyle{ (3 N - 4)!^2 \over ( 2 N - 3 )!^3 } \,\,
\displaystyle{ N! \over 2^N } \,\,
\left\langle \displaystyle{1\over C^2} \right\rangle
\end{equation}
where
\begin{equation}
\left\langle \displaystyle{1\over C^2} \right\rangle
\equiv
\int \, dC \, \displaystyle{ p_c \over C^2 } \,\, .
\end{equation}
Using the H\"{o}lder inequality
to show that $\langle 1/C^2 \rangle \ge 1/\langle C \rangle^2,$
one has
\begin{equation}
\label{eq:sumi}
\sum_i I_i \,\, \ge \,\,
\displaystyle{ (3 N - 4)!^2 \over ( 2 N - 3 )!^3 }  \,\,
\displaystyle{N! \over 2^N } \,\,
\displaystyle{1\over \langle C \rangle^2 }  \,\, .
\end{equation}

We have studied the complexity density functions
of $K-$graphs up to $N=400$ and we find (see Appendix~A)
the empirical asymptotic relation
\begin{equation}
\label{eq:empirical}
\langle C \rangle \simeq
\displaystyle{ .56 \over  N }
\left( \displaystyle{ 27 \over 8 } \right)^N
\end{equation}
which works to better than $\simeq 5\%$ for $N \ge 20.$
(In the range $20 \le N \le 400, $
$\langle C \rangle$ varies by $200$ orders of magnitude,
so this is a very accurate fit for the constant $27/8$ in
Eq.~\ref{eq:empirical}.)
Using Eq.~\ref{eq:empirical} in Eq.~\ref{eq:sumi}
we find, for $N \gg 1, $
\begin{equation}
\sum_i I_i \,\, \ge \,\,
\displaystyle{ 1  \over (.56)^2 }
\left( \displaystyle{ 2 \over 3 }\right)^7
\displaystyle{ N^2 \over \sqrt{2} }
\left( \displaystyle{ 1 \over 2 } \right)^N
N!
\doteq
\left( \displaystyle{ 1 \over 2 } \right)^N
N!
\end{equation}
In other words, the contribution of all $N^{\rm th}$ order
$K-$graphs to $Im~T_{2 \rightarrow 2}$ is
\begin{equation}
\label{eq:recent}
Im~T_{2 \rightarrow 2 } \doteq
 N! \left( \displaystyle{ a_K g \over 16 \pi^2 } \right)^N,
\end{equation}
where $ a_K \ge 1/2.$  If one includes all $g \phi^4$ graphs without
two-line loops, the corresponding limit is $a \ge 2/3$ (see Appendix E).
These bounds may also be derived without using the complexity
results of Eq.~\ref{eq:empirical} by instead bounding the
asymptotic behavior of the integrals
$\int [dx] / U_i^2 \ge \int [dx] / U_S^2$ where $U_S$ is
is the completely symmetric sum of products of $N-1$
Feynman parameters (see Appendix B). In Appendix~B we
also use integrals of $U_S$ to speculate upon possible
improvements to the bounds on $a.$ These speculations amount
to multiplying $a_K$  by the ratio $C_S^2/\langle C\rangle^2,$
where $C_S \doteq 4^N $ is the complexity of $U_S;$
this leads to estimating $a_K \geq 512/729 \simeq 0.7$ if
one includes only $K-$graphs and $a \geq 2048/2187 \simeq 0.94$
if one includes all simple graphs as in Appendix~E.
Presumably the answer for the sum of all graphs without
logarithms is Eq.~\ref{eq:recent} with $a=1,$ the Lipatov result.
Though $K-$graphs are fundamentally rooted in Minkowski space
at large momenta they have much in common with graphs having no
external momenta, as required by the Lipatov technique.
In Appendix C we point out striking statistical regularities
which arise in the integrals of the $U_i.$

\section{Summing $K-$Graphs and Two-Line Loops}

\hspace*{\parindent}
In this section we investigate the marriage of
the factorial behavior of $K-$graphs with
leading logarithmic behavior in $g \phi^4$ theory.
Specifically, we sum contributions to $Im~T_{2 \rightarrow 2}$
from chain graphs of the form
shown in Fig.~\ref{fi:bssol} which are constructed
by joining arbitrary combinations of $K-$graphs and
two-line loops. The sum of all such graphs
and their $u-$channel exchange graphs is summarized by
the Bethe-Salpeter equation depicted in Fig.~\ref{fi:bseq}.

A key result which makes summing all chain graphs feasible
is a factorization theorem illustrated by Fig.~\ref{fi:factor};
we prove the theorem in Appendix~\ref{se:proof}.
The essence of the theorem is that, up to leading logarithms,
one may factor a  $K-$graph from a graph and
replace it with a two-line loop. The constant of
proportionality between the imaginary parts of the two graphs is
\begin{equation}
\label{eq:khatdef}
\hat{K} = \displaystyle{
\left( Im~T_{2 \rightarrow 2}\right)_K \over
\left( Im~T_{2 \rightarrow 2} \right)_{N=2} } =
2 \left( \displaystyle{ g \over 16 \pi^2 } \right)^{N_K-2}
\int_0^1 \displaystyle{ [dx] \over U^2_K }\, \,,
\end{equation}
where $\left( Im~T_{2 \rightarrow 2}\right)_K$ is the contribution to
$Im~T_{2 \rightarrow 2 }$ from the corresponding
isolated $N^{\rm th}_K$
order $K-$graph and its $u-$channel exchange graph
(e.g., as in Fig.~\ref{fi:bseq}d,e) and
$\left( Im~T_{2 \rightarrow 2} \right)_{N=2}$ is the contribution
from a single two line loop (e.g., as in Fig.~\ref{fi:bseq}b,c).

The ability to factor $K-$graphs out of chain graphs
effectively reduces the problem of solving the Bethe-Salpeter
equation of Fig.~\ref{fi:bseq} to the much simpler and
well-studied\cite{sawyer,arb} problem of summing ladder graphs.
Denoting terms in the Bethe-Salpeter equation
by their labels in Fig.~\ref{fi:bseq}, one has from
Fig.~\ref{fi:factor} and Eq.~\ref{eq:khatdef},
\begin{eqnarray}
( Im~T_{2 \rightarrow 2})_b + (Im~T_{2 \rightarrow 2})_c & = &
( Im~T_{2 \rightarrow 2})_{N=2} \,\, , \\
(Im~T_{2 \rightarrow 2})_d + (Im~T_{2 \rightarrow 2})_e & = &
 \left(  \sum \hat{K}_i \right) \,
(Im~T_{2 \rightarrow 2})_{N=2} \,\,,
\\
(Im~T_{2 \rightarrow 2})_f + (Im~T_{2 \rightarrow 2})_g & = &
\left( 1 + \sum \hat{K}_i \right) \,(Im~T_{2 \rightarrow 2})_f \,\, ,
\end{eqnarray}
so that the Bethe-Salpeter of Fig.~\ref{fi:bseq} becomes
\begin{equation}
\label{eq:newbs}
(Im~T_{2 \rightarrow 2})_a =
\left( 1 + \sum \hat{K}_i \right)
\left[ (Im~T_{2 \rightarrow 2})_{N=2} + (Im~T_{2 \rightarrow 2})_f \right] \, .
\end{equation}
The terms in square brackets of Eq.~\ref{eq:newbs} are recognized
as the right hand side of the Bethe-Salpeter equation appropriate
for summing straight ladders.
Consequently, the Bethe-Salpeter equation involving
$K-$graphs is solved by taking the solution to the
corresponding sum of ladder graphs (Eq.~\ref{eq:arb})
and making the substitution
\begin{equation}
\label{eq:gsub}
g \rightarrow g \left( 1 + \sum \hat{K}_i \right)^{1/2} \,\, .
\end{equation}
With the substitution of Eq.~\ref{eq:gsub} we obtain
\begin{equation}
Im~T_{2\rightarrow2}( s \gg m^2 , t = 0 ) \sim
\displaystyle{g^2 \left( 1 + \sum \hat{K}_i \right)
\over \ln^{3/2}\left( s/m^2 \right)  }
\left[ \left( \frac{s}{m^2} \right)^{ \frac{g}{16 \pi^2}
\left( 1 + \sum \hat{K}_i \right)^{1/2} }
+ ( g \rightarrow -g )  \right]
\end{equation}
where, from section~\ref{se:kgraphs},
\begin{equation}
\sum \hat{K}_i  \doteq
\sum_{\rm{even}~N} N! \left( \displaystyle{  a g \over 16 \pi^2 } \right)^N .
\end{equation}

Naturally, Eq.~\ref{eq:gsub} can also be deduced
by considering the effect of the factorization theorem
on individual chain graphs like Fig.~\ref{fi:bssol}. Consider the set of
chain graphs with $N/2$ ``rungs'' where a rung is either a
$K-$graph or a two-line loop. By the factorization theorem,
the sum of all such chains is proportional to the
$N^{\rm th}$ order uncrossed ladder graph (where all
rungs are two-line loops) and the factor of proportionality
follows by summing over the possible replacements of two-line
loops with $K-$graphs. Let $n_i$ denote the number of
times a $K_i-$graph appears in the chain. The
factorization theorem and elementary
combinatorics give the overall
factor of proportionality
\begin{equation}
\label{eq:multinom}
\sum_{\{n_i\}}
\displaystyle{ (N/2)!
\over
n_{i_1}! n_{i_2}! \cdots
\left( N/2 - \sum_i n_i \right)! }
\prod_i \hat{K}_i^{n_i} \, ,
\end{equation}
where the sum over ${n_i}$ is such that
$\sum_i n_i \le N/2.$ Eq.~\ref{eq:multinom} is
simply the multinomial expansion of
$\left( 1 + \sum \hat{K}_i \right)^{N/2}$ so that summing
over all insertions of $K-$graphs is equivalent to
the ladder graph result with the substitution of Eq.~\ref{eq:gsub}.

\section{Conclusions}

\hspace*{\parindent}
By summing graphs in $g \phi^4$  theory,
 we have shown how a Regge-like energy behavior
and a Lipatov-like $N! g^N$ behavior
co-exist at high energies. The Lipatov
analysis is restricted to zero four-momentum, and fails to
be directly applicable to overall high-energy process. But
we have found classes of graphs (like $K-$graphs)
which very much resemble zero-momentum graphs and which
give various hints of being amenable to
semi-classical analysis. These graphs factor out of
energy-dependent processes and can be analyzed separately.
Presently we do not know how to calculate sets of graphs
like $K-$graphs with semi-classical techniques, but there surely must be
such a method which might ultimately tell us how to factor all relevant
graphs out of high-energy processes and how to calculate the
coefficient $a$ in the expansion
$N! \left(  \frac{a g }{16 \pi^2} \right)^2$
without approximation. (Of course, we expect to find the Lipatov value
$a=1).$

Many other issues remain unsettled. For example,
our graphical analysis occurred in the absence
of derivative couplings and involved the dimensionless
quantities $g$ and $Im~T_{2 \rightarrow 2};$ these are
the circumstances under which one also expects
to succeed with a Lipatov analysis. But in realistic theories like QCD,
and even in $g\phi^4$ (e.g., if studying $T_{2\rightarrow N})$
one or all of these conditions is not met.

	Consider, for example, the amplitude $T_{2 \rightarrow N}$
in $g \phi^4$ theory. The very special case where
all but two of the final-state particles have zero four-momentum
can be obtained from our previous results.
In the basic graph of Fig.~\ref{fi:tree} from which all our other
graphs are derived, let the masses of the
left-hand vertical lines be $M_1$ and that of the right-hand
lines be $M_2;$ roughly speaking, this replaces $m^2$ in
$\ln (s/m^2)$ by $\frac{1}{2}(M_1^2 + M_2^2).$
Applying $g \frac{\partial }{\partial M_2^2}$
$N$ times to $Im~T_{2 \rightarrow 2}$ produces a
contribution to $Im~T_{2 \rightarrow 2 N + 2 }$
whose energy dependence is unchanged from that of
$Im~T_{2 \rightarrow 2};$ it has simply picked up a factor
of $M_2^{-2 N}.$ Yet when the $2N$ added particles all have large momenta,
of order $\frac{\sqrt{s}}{2N},$ we expect an energy dependence where
$M_2^{-2N}$ is replaced by $s^{-N}.$ This regime, when
many particles have large energy, simply cannot be addressed in any
straightforward way by Lipatov techniques. The difficulty also
shows up in the graphical analysis  where the relatively simple
integrals for $K-$graphs such as Eq.~\ref{eq:imt} are replaced
by integrals which involve powers of the energy-dependent function $\phi.$
Even more drastic changes occur for QCD where, because of derivative
couplings, not only do powers of $\phi$ appear but altogether different
combinations of Feynman parameters arise\cite{cheng2}.

	There is another stumbling block on the road to
QCD. At high energy, with $s \gg |t|,$ it is
well-known\cite{fic} that QCD Reggizes somewhat as
$\phi^4$ theory does, but the equivalent of the
$K-$graphs are not four-dimensional Feynman graphs,
as in $\phi^4$ theory, but two-dimensional,
referring only to momentum transverse to $p$ and $p'.$ The
underlying two-dimensional theory is presumably a sigma
model, as the Verlindes\cite{verlind} have suggested, but the
usual Lipatov analysis has not been applied here.

	We do, however, know something about leading-logarithm
behavior in QCD when $s\gg t\gg m^2\cite{fic}$ where $m^2$ is either
a fictitious mass or a Higgs mass for the gauge boson\cite{corn4}. The
analog of Eq.~\ref{eq:arb} for non-forward scattering in QCD yields
amplitudes which vary as $s^{\alpha_P(t)}$ where the ``Pomeron'' trajectory
$\alpha_P$ is given by, to $O(\alpha_S)$ in the strong coupling
constant $\alpha_S,$
\begin{equation}
\label{eq:pomeron}
\alpha_P(| t | \gg m^2 ) = 1 - \displaystyle{\alpha_S N_C \over 2 \pi }
\ln\left(\displaystyle{ |t| \over m^2 }\right),
\end{equation}
where $N_C$ is the number of colors. Suppose, for the sake of argument,
that including the QCD analog of $K-$graphs leads to the replacement
\begin{equation}
\label{eq:qcdreplace}
\alpha_S \rightarrow \alpha_S \left(
1  + \sum_N
\left[
- \displaystyle{\alpha_S N_C \over 2 \pi }
\ln\left(\displaystyle{ |t| \over m^2 }\right)
\right]^N N! \right)
\,\, .
\end{equation}
Eq.~\ref{eq:qcdreplace} corresponds to Eq.~\ref{eq:gsub1} except that
now there is a $t-$dependence due to non-forward scattering.
At this point, only leading IR logs have been kept but it is possible
to renormalization group (RG) improve Eq.~\ref{eq:pomeron}
(and, correspondingly Eq.~\ref{eq:qcdreplace})
and incorporate the effects of the one-loop running charge
\begin{equation}
\bar{\alpha}_S(t) = \displaystyle{ 1 \over 4 \pi b \ln(|t|/m^2) }\, ,
\qquad
b = \displaystyle{ 1 \over 48 \pi^2} ( 11 N_C - 2 N_F )  \,  ,
\end{equation}
where $N_F$ is the number of flavors.
The RG-improved result is not quite the naive result of replacing
$\alpha_S$ by $\bar{\alpha}_S$\cite{derujula} but rather,
Eq.~\ref{eq:qcdreplace} becomes
\begin{equation}
\label{eq:last}
\bar{\alpha}_S
\rightarrow
\bar{\alpha}_S
\left(1+
\sum_N \gamma^N(t) N!\right) ,
\qquad
\gamma(t) = \displaystyle{ 6 N_C\over
11 N_C - 2 N_F }
\ln \left(  \ln\left(\displaystyle{ |t| \over m^2 } \right) \right).
\end{equation}

Previous experience\cite{corn4} with unitarizing factorially divergent
series such as Eq.~\ref{eq:last} suggests that the largest value
of $N$ which can be trusted is $N = 1/\gamma,$ but this is typically
not even as large as 2. QCD perturbation theory for high-energy fixed-$t$
processes is subject to large effects from $N!$ divergences
even for small $N.$ This is not what are might
have guessed by looking at the kind of
series one gets just from the Lipatov analysis,
which applies to purely $s-$wave processes and looks something
like
\begin{equation}
\sum_N \left(
            \displaystyle{ \alpha_S \over 2 \pi}
     \right)^N
     N!
\qquad {\rm or}
\qquad
\sum_N
      \left( \displaystyle{ \bar{\alpha}_S \over 2 \pi} \right)^N N!
\end{equation}
The second form is, of course, RG-improved, with the argument
of $\bar{\alpha}_S$ now being $s,$ not $t.$ The RG-improved
series begins to diverge at
$N \simeq \frac{1}{2} ( 11 N_C - 2 N_F ) \ln(s /m^2)$
which typically is considerably larger than one.

\section{Acknowledgements}

JMC is supported by the National Science Foundation under
grant PHY-9218990. We would like to thank D.~Cangemi
and S. Nussinov for many discussions.

\appendix
\section{Complexity of $K-$Graphs}

\hspace*{\parindent}
In this appendix we present results on the
distribution of complexity of $K-$graphs.
A useful result from algebraic graph theory is an expression for
the complexity of an $N^{\rm th}$ order graph\cite{biggs,temperley},
\begin{equation}
C = \displaystyle{ \det( J + Q ) \over N^2},
\end{equation}
where $J$ is an $N \times N$ matrix whose matrix elements are all
equal to unity, and $Q$ is a $N \times N$ symmetric matrix.
For $2\rightarrow 2$ graphs in $g \phi^4$ theory, $Q$
may be constructed as follows. Draw an $N^{\rm th}$ order graph with
its external lines truncated and label the vertices from $1$ to $N.$
The diagonal matrix elements $Q_{ii}$ are the number of lines
attached to the $i^{\rm th}$ vertex. The off diagonal matrix
elements are such that $-Q_{ij}$ is the number of lines joining vertex
$i$ to vertex $j.$

Figure~\ref{fi:comp} shows the distribution of complexities for all
43,200 $K-$graphs for $N=12.$ Table~1 lists the average
complexity of $K-$graphs for various orders of perturbation theory.
A least-squares fit of $\langle C \rangle$ over
$20 \le N \le 400$ to the form $c_1 N^{c_2} c_3^N$
(with $c_2$ an integer) yields the best fit
\begin{equation}
\langle C \rangle \simeq {.5585 \over N} (3.3752)^N ,
\end{equation}
which is accurate to better than $0.5\%$ (comparable
to the accuracy with which the actual averages are determined).
In fact, using $ \langle C \rangle \simeq {.5585 \over N} (27/8)^N $
works better than $5\%$ which is impressive since $\langle C \rangle$
varies by 200 orders of magnitude over the range in question.
Though we have not pursued an analytic calculation of
$\langle C \rangle$ for $K-$graphs, it not unreasonable to
believe that such an approach exists for asymptotically large $N$
since analogous graph-theoretic problems have been
encountered previously in the study of cluster integrals
in statistical mechanics\cite{temperley}. The near-Gaussian
nature of the complexity distribution and the $N-$dependence
of $\langle C \rangle$ are analogous to the observations of
Bender and Wu\cite{ben76} concerning the behavior of connected
vacuum diagrams in $\phi^4$ theory in one spacetime dimension.

\section{Bounds From Completely Symmetric $U_S$}

\hspace*{\parindent}
The functions $U_i$ characterizing contributions of $N^{\rm th}$
order $K-$graphs to $Im~T_{2 \rightarrow 2}$ share the feature that
they are sums of monomials formed from the products of
$N-1$ Feynman parameters (out of a possible $2 N - 2$ parameters).
Since the number of monomials in $U_i$ (that is, the complexity $C_i$)
varies from graph to graph it is in practice difficult, if not
impossible, to obtain analytic expressions for integrals of the form
$\int [dx] / U_i^2 .$ However, one can hope to evaluate
integrals of the completely symmetric function $U_S$ which, at
$N^{\rm th}$ order, is the symmetric sum of all possible
combinations of products of $N-1$ parameters chosen from $2 N - 2$
parameters. Since $U_S \ge U_i,$ one obtains the
bound $\int_0^1 [dx] / U_i^2  \ge \int_0^1 [dx] / U_S^2.$

In this appendix we investigate the large-$N$ behavior of the integral
\begin{equation}
\label{eq:sumk3}
I_S(2 N - 2)\equiv \int_0^1 \frac{ [ dx ] }{ U_S^2 },   \qquad \qquad
[ dx ]  =
dx_1 dx_2 \cdots dx_{2 N - 2} \, \,
\delta \! \left( 1 - \sum_j^{2N-2} x_j \right)
\end{equation}
and we use our results to place lower bounds on the corresponding
integrals of the $U_i$ functions. To our knowledge $I_S(2 N -2)$
cannot be performed directly so instead, letting $ n = 2 N - 2, $
we concentrate on
\begin{equation}
\label{eq:basic}
I_S(n \gg 1 ,k) \equiv \int [dx] \, U_S^k
\end{equation}
for non-negative integer $k$ with the intention of
analytically continuing to $k=-2.$
In Sect.~\ref{se:appu1} we find that,
to leading order in $n,$
\begin{equation}
\label{eq:result}
I_S(n \gg 1,k) \simeq
\displaystyle{ ( 2 \pi n )^{-k/2} \over
\Gamma \left( n \left(
\displaystyle{ k \over 2 } + 1  \right)\right)
}
\left( \displaystyle{ 2 z^{-k/2}  \over 2 - kz } \right)^n
\sqrt{
\left(
\displaystyle{  2 z ( 1 - k ) \over z-1 } \right)^{k-1}
\displaystyle{  4 z           \over z ( k + 2 ) - 2  } } \,\, ,
\end{equation}
where $z$ is a function of $k$ given implicitly by
\begin{equation}
\label{eq:zk}
\displaystyle{ d \over dz}
\left(
z^{k/2} e^{1/z} \Gamma(k+1,1/z)
\right) = 0 ,
\end{equation}
where $\Gamma(k+1,1/z) = \int^\infty_{1/z}  dt \, e^{-t} t^k$ is the
incomplete gamma function.

In Sect.~\ref{se:appu2} we demonstrate that
Eq.~\ref{eq:result} yields impressive agreement
with numerical integration of $I_S(n,k)$
for $ k \gwig -2$ but becomes
unreliable very close to $k=-2$ (when $|k+2| \lwig O(1/n)).$
Nevertheless, we argue that the large-$n$ behavior of $I_S(n,k=-2)$
is already contained in Eq.~\ref{eq:result} and that,
modulo an overall constant and a fixed power of $n,$
$I_S(n \gg 1 ,k=-2) \doteq 1.$

In Sect.~\ref{se:appu3} we demonstrate how the
bound $Im~T_{2 \rightarrow 2} \doteq
N! \left(\frac{ a g }{16 \pi^2}\right)^N$ with
$a_K \ge 1/2$ follows from a
consideration of symmetric functions. In addition,
we speculate about upon possible improvements to
which may make $a_K \simeq 0.7.$ Analogous reasoning
is applied to the set of all simple graphs (of which
$K-$graphs are a subset) in Appendix~E.

\subsection{Saddle Point Evaluation of $I(n \gg 1,k)$}
\label{se:appu1}

\hspace*{\parindent}
For non-negative integer $k$ one can expand $U_S^k$ and
integrate term by term to get
\begin{equation}
\label{eq:compactsum}
I_S(n,k) = \displaystyle{ n! \over ( n + kn/2 - 1)! }
\sum_{  \{  n_j^{(q)} \} }
\left( \prod_{j,(q)}
\displaystyle{ (j!)^{n_j^{(q)}} \over n_j^{(q)}!
 }\right),
\end{equation}
where the sum over the non-negative integer variables
$\{ n_j^{(q)} \}$ is subject to the $k+1$ constraints
\begin{equation}
\sum_{j=0}^{j=k} \sum_{(q)} n_j^{(q)}    =  n , \qquad \qquad
\sum_{j=1}^{j=k} \sum_{(m \, q)} n_j^{(m \, q)}  =  \frac{n}{2}
\qquad\qquad {\rm for}~m = 1,\cdots,k.
\end{equation}
We will not discuss the motivation for introducing the
$ \{ n_j^{(q)} \}$ except to point out that they arise
naturally from a combinatoric analysis
of terms in the expansion of $U_S^k.$
Our notation for ${ n_j^{(q)} }$ is such that $(q)$ is
a combination of $j$ integers chosen from 1 to $k.$
For example, for $k=3$ the relevant set of variables is
$\{ n_0^{(0)},
n_1^{(1)},
n_1^{(2)},
n_1^{(3)},
n_2^{(12)},
n_2^{(13)},
n_2^{(23)},
n_3^{(123)} \}; $ in general there are $2^k$
different $n_j^{(q)}.$ The label $( m \, q)$  in $n_j^{(m \, q )}$
denotes a combination of $j$ integers which includes the integer $m.$

To find $I_S(n \gg 1,k)$ we replace the discrete sum
in Eq.~\ref{eq:compactsum} with an integral of the
corresponding continuous quantity
\begin{equation}
\label{eq:inter}
I_S(n\gg 1,k) \simeq
\displaystyle{
\Gamma(n+1) \over
\Gamma \left( n \left( \frac{k}{2} + 1 \right) \right) }
\int
\prod_{j,(q)}
\displaystyle{ dn_j^{(q)} (j!)^{ n_j^{(q)} } \over \Gamma( n_j^{(q)}+1)
}
\prod_{m=0}^{m=k} \frac{ d\lambda_m }{ 2 \pi }
e^{i \lambda_m f_m } \, ,
\end{equation}
where the constraints have been incorporated by defining
\begin{equation}
\label{eq:compactconst}
f_0 = n \, - \, \sum_{j=0}^{j=k} \sum_{(q)} n_j^{(q)},  \qquad
f_{m} = \frac{n}{2}
\, - \, \sum_{j=1}^{j=k} \sum_{(m \, q)} n_j^{(m \, q)},
\qquad {\rm for}~m = 1,\cdots,k  .
\end{equation}
With the approximation $\Gamma(p+1) \simeq \sqrt{2 \pi p}\, (p/e)^p$
we can rewrite Eq.~\ref{eq:inter} in the form
\begin{equation}
\label{eq:messyintegral}
I_S(n\gg 1,k) \simeq
\displaystyle{ \Gamma(n+1) \over
\Gamma \left( n \left( \frac{k}{2} + 1 \right) \right)
(2 \pi)^{(k+1+ 2^{k-1})} }
 \int
\prod_{j,(q)}  dn_j^{(q)} \prod_{m=0}^{m=k} d\lambda_m
e^F ,
\end{equation}
where the argument of the exponential is
\begin{equation}
F = \sum_{j=0}^{j=k}  \sum_{(q)} \left(
n_j^{(q)} \ln j! - ( n_j^{(q)} + \frac{1}{2} )
\ln n_j^{(q)} + n_j^{(q)} \right) +
i \sum_{j=0}^{j=k} \lambda_j f_j \, \, .
\end{equation}
Using Laplace's method to evaluate $I_S(n \gg 1, k)$ we
find that the extremum of $F$ occurs when
$n_j^{(q)} = \bar{n}_j \equiv j! x z^j $ where
$z \equiv e^{i \bar{\lambda} }, x \equiv
e^{ i \bar{\lambda}_0}$ and overbars denote parameter
values at the extremum.

At the extremum, $f_0=f_m=0$ so that the
constraints of Eq.~\ref{eq:compactconst} become
\begin{equation}
x \sum_{j=0}^{j=k}
\left( \begin{array}{c} k \\ j \end{array} \right)
 j! z^j  = n , \qquad \qquad
x \sum_{j=1}^{j=k}
\left( \begin{array}{c} k-1 \\ j-1 \end{array} \right)
j! z^j  = \frac{n}{2},
\end{equation}
which may be rewritten as
\begin{equation}
\displaystyle{ d \over dz}
\left(
z^{k/2} e^{1/z} \Gamma(k+1,1/z)
\right) = 0, \qquad \qquad
x =   \displaystyle{ n \over 2 } ( 2 - k z ) .
\end{equation}

We can proceed to perform the integrations in
Eq.~\ref{eq:messyintegral} by expanding $F$ around its maximum
\begin{equation}
F = \left. F \right|_{\bar y} +
\sum_{i,j = 1}^{2^k + k + 1}
\displaystyle{1\over 2}
\left.
\displaystyle{ \partial^2 F \over \partial y_i \partial y_j }
\right|_{\bar{y}}
( y_i - \bar{y}_i ) ( y_j - \bar{y}_j ) + \cdots
\end{equation}
where the first $2^k$ of the $y_i$ variables are the
$n_j^{(q)}$ and the remaining $k+1$ variables
are  $i \lambda_j$ for $(j=0, ..., k).$
Performing the Gaussian integrations we obtain
\begin{equation}
\label{eq:integrated}
I_S(n\gg 1,k) \simeq
\displaystyle{ ( 2 \pi )^{-k/2} \over
\Gamma \left(  n \left( \frac{k}{2} + 1 \right) \right) }
\left( \displaystyle{ 2 z^{-k/2} \over  2 - k z } \right)^n
\sqrt{
\displaystyle{ n \over {\rm det}(A) }
\prod_{j(q)} \displaystyle{ 1 \over \bar{n}_j^{(q)} } } .
\end{equation}
where $A$ is the matrix
defined by
\begin{equation}
A_{ij} =
- \left.
\displaystyle{ \partial^2 F \over \partial y_i \partial y_j }
\right|_{\bar{y}} .
\end{equation}

After a somewhat lengthy but straightforward manipulation,
det(A) can be written as
\begin{equation}
{\rm det}(A)
= (-n)^{k+1}
\displaystyle{ 2 z + k z - 2   \over 4 z }
\left(
\displaystyle{ z- 1 \over  2z ( 1- k ) } \right)^{k-1}
\prod_{j (q)}
\displaystyle{ 1 \over \bar{n}_j^{(q)}} ,
\end{equation}
which, when substituted in Eq.~\ref{eq:integrated} gives
the result of Eq.~\ref{eq:result}.

\subsection{The Limit $k\rightarrow -2$}

\label{se:appu2}
\hspace*{\parindent}

We have compared $I_S(n \gg 1,k)$ of Eq.~\ref{eq:result}
with results from integrating
$\int [dx] U_S^k$ numerically for $n \le 20.$ As can be seen from
Fig.~\ref{fi:asym}, where we show the comparison for $n=20,$
the analytic continuation of Eq.~\ref{eq:result}
to negative $k$ is successful (agreeing up to a
correction factor of $1 + O(1/n)$) until $k \simeq  -2.$
The nature of the discrepancy is made evident
by noting that as $k \rightarrow -2,$ Eq.~\ref{eq:zk}
reduces to the asymptotic relation
\begin{equation}
 k + 2 \simeq \frac{2 \ln z}{z} \,\, ,
\end{equation}
As $k\rightarrow 2,$  Eq.~\ref{eq:result} becomes
\begin{equation}
\label{eq:la1}
I_S( n \gg 1, k \simeq -2 ) \simeq \frac{\pi n^2 }{3 \sqrt{6}} \, \sqrt{k+2}.
\end{equation}
which suggests that the neglected corrections are such that
\begin{equation}
\label{eq:cor}
I_S( n \gg 1, k \simeq -2 ) \simeq
\frac{\pi n^2 }{3 \sqrt{6}} \, \sqrt{k+2} \,
\left[ 1 + O\left( \frac{1}{n \sqrt{k+2} } \right) \right].
\end{equation}
Indeed, numerical integration of $I_S(n,k=-2)$ for $6 \le n \le 20$
is consistent with $I_S(n,k=-2) \sim n.$

In any case, our goal is to extract the
large$-N$ behavior of $I_S(2 N - 2, k = -2).$  As has already been
exploited in taking the $k \rightarrow -2$ limit of
Eq.~\ref{eq:result} to arrive at Eq.~\ref{eq:la1},
\begin{equation}
\lim_{k \rightarrow -2}
\left( \displaystyle{ 2 z^{-k/2}  \over 2 - kz } \right)^n = 1^n \,\, ,
\end{equation}
since $z\rightarrow \infty$ as $k \rightarrow -2.$
In other words, with the $O(1/N)$ corrections of
Eq.~\ref{eq:cor} taken into account,
\begin{equation}
\label{eq:central}
I_S(2 N - 2 \gg 1, k= -2) \doteq 1.
\end{equation}

\subsection{Bounds on $Im~T_{2 \rightarrow 2}$}
\label{se:appu3}

\hspace*{\parindent}
Without making recourse to the complexity results of Appendix~A one can
reproduce the lower bound $a_K \ge 1/2$ for contributions to $Im~T_{2
\rightarrow 2}$ simply by considering integrals of the completely symmetric
functions $U_S$. Since $U_S \ge U_i,$ it follows that the contributions
of all $N^{\rm th}$ order $K-$graphs obey the bound
\begin{eqnarray}
\label{eq:simplebounds}
Im~T_{2 \rightarrow 2} & = &
\displaystyle{ g^2 \over 16 \pi }
\left( g \over 16 \pi^2 \right)^{N-2} \sum_i
\int \displaystyle{ [dx] \over U_i^2 }  \nonumber \\
  & \ge &
\displaystyle{ g^2 \over 16 \pi }
\left( g \over 16 \pi^2 \right)^{N-2} \sum_i
\int \displaystyle{ [dx] \over U_S^2 } \nonumber \\
 &  \doteq &
\displaystyle{ N! \over 2^N }
\left( g \over 16 \pi^2 \right)^{N}
\nonumber \\
 &  = &
N! \left( a g \over 16 \pi^2 \right)^N
\end{eqnarray}
with $a = 1/2.$  In going from the second to third lines
of Eq.~\ref{eq:simplebounds}, we have used the asymptotic
behavior of Eq.~\ref{eq:central} and summed over the number
of $N^{\rm th}$ order $K-$graphs which is $\doteq N!/2^N.$

If we wish to be adventuresome we can contemplate
extending the lower bounds on $Im~T_{2 \rightarrow 2}$
even further by combining the complexity results of
Appendix~A with the asymptotic behavior of the integrals of $U_S.$
Letting $C_S = (2 N - 2)! / (N-1)!^2 \doteq 4^N $
denote the number of terms in $U_S,$
it follows from Eq.~\ref{eq:averages} that
\begin{equation}
\label{eq:av}
\displaystyle{ \langle U_i \rangle \over C_i } =
\displaystyle{ \langle U_S \rangle \over C_S } ,
\end{equation}
where $C_i$ is the complexity of $U_i.$
Although Eq.~\ref{eq:av} strictly only refers to the average
values of $U_S$ and $U_i,$ it is amusing to explore the
consequences of extending the relation $U_i \le U_S$ by assuming
that, in the region $U_i \simeq 0$ (where the important
contributions to $\int [dx] / U_i^2 $  arise),
\begin{equation}
\label{eq:assumption}
\displaystyle{ U_i \over C_i }
\,\, \lwig \,\,
\displaystyle{ U_S \over C_S } .
\end{equation}
The motivation for this assumption is that when $C_i$ becomes large,
then for fixed values of the $2 N - 2$ Feynman parameters
$\{x_i\},$ $U_i/C_i$ can be thought of as an estimate of the average value
of all $C_S$ possible monomials formed from combinations
of $N-1$ of the $x_i;$ the true average of these monomials
(for the same fixed $\{ x_i \})$ is, naturally,  $U_S/C_S.$
Of course, the monomials in $U_i$ are not random --- they arise
from $K-$graphs. The assumption that $U_i/C_i$ is smaller than $U_S/C_S$
is an attempt to reflect the correlations among monomials in $U_i$
make which it easy for $U_i$ to be small: only 3 Feynman
parameters (corresponding to three-line loops)
need vanish simultaneously for $U_i$ to vanish whereas
$N-1$ parameters must simultaneously vanish for $U_S$ to go to zero.

Under the assumption of the inequality of Eq.~\ref{eq:assumption}
the contribution of all $N^{\rm th}$ order $K-$graphs is
\begin{eqnarray}
\label{eq:improve}
Im~T_{2 \rightarrow 2} & = &
\displaystyle{ g^2 \over 16 \pi }
\left( g \over 16 \pi^2 \right)^{N-2} \sum_i
\int \displaystyle{ [dx] \over U_i^2 }  \nonumber \\
& \ge &
\displaystyle{ g^2 \over 16 \pi }
\left( g \over 16 \pi^2 \right)^{N-2} \sum_i
\left( \displaystyle{ C_S \over C_i } \right)^2
\int \displaystyle{ [dx] \over U_S^2 } \nonumber \\
&  \doteq &
\left( g \over 16 \pi^2 \right)^{N} \sum_i
\left( \displaystyle{ 4^{N} \over ( 27 / 8 )^N  } \right)^2
\nonumber  \\
&  \doteq &
N! \left( a_K g \over 16 \pi^2 \right)^N
\end{eqnarray}
with $ a = 512/729 \simeq .70$
and where we have used the asymptotic behavior of $I_S$ and the
empirical average complexity of the $K-$graphs found in Appendix~A.
We emphasize that the inequality of Eq.~\ref{eq:assumption},
though plausible, is only an assumption. It is consistent with our
numerical work where limitations of computer time make checks
feasible. It is plausible that the bound $a_K \ge 1/2$
or perhaps even $a_K \simeq 0.7$ could be sharpened
with a judicious application of inequalities to the information
already at hand.

\section{Regularity of Contributions To $Im~T_{2 \rightarrow 2}$}

\hspace*{\parindent}
In this appendix we present evidence for regularities in the
contributions of $N^{\rm th}$ order $K-$graphs to
$Im~T_{2 \rightarrow 2}.$ The sum of these contributions
(including $u-$channel exchange graphs) is
\begin{equation}
\label{eq:imagain}
Im~T_{2 \rightarrow 2}  =
\displaystyle{ g^2 \over 16 \pi }
\left( g \over 16 \pi^2 \right)^{N-2} \sum_i
\int_0^1 \displaystyle{ [dx] \over U_i^2 } \,\, ,
\end{equation}
where $U_i$ characterizes the $i^{\rm th}$ $K-$graph.
In section \ref{se:kgraphs} we defined for each graph the
probability density $p_i$ which treats $U$ as an independent
variable and we derived lower bounds for the integrals of
Eq.~\ref{eq:imagain} in terms of the expectation values
$\langle U \rangle_i = \int dU p_i U $ (see Eq.~\ref{eq:lowb}).
Whereas the distribution of the complexity $C_i$
(see e.g., Fig.~\ref{fi:comp}) reflects regularity
in $\langle U \rangle_i,$ we now wish to
point out a strong correlation between the shapes
of the probability densities $p_i$ themselves.

To facilitate a comparison of the $p_i,$ we
first rescale from the variable $U$ to the variable
$\tilde{U} = U / \langle U \rangle_i$
so that $\langle \tilde{U} \rangle_i = 1$ for all $K-$graphs.
In other words, it is convenient to compare the
probability densities $\tilde{p}_i$ defined by
\begin{equation}
\tilde{p}_i =
\displaystyle{\int_0^1 [dx]
\delta\!\left( \tilde{U} -
\displaystyle{ U_i(x) ( 3 N - 4 )!\over C_i}
\right) \over \int_0^1 [dx] }.
\end{equation}
In terms of the $\tilde{p}_i,$
\begin{equation}
\int_0^1 \displaystyle{ [dx] \over U_i^2 }
= \displaystyle{ ( 3 N - 4 )!^2 \over ( 2 N - 3 )! }
 \displaystyle{ 1 \over C_i^2 }
\int d\tilde U \, \, \displaystyle{ \tilde{p}_i \over \tilde{U}^2 }.
\end{equation}

Figure~\ref{fi:72} superimposes the results of Monte Carlo
calculations of $\tilde{p}_i$ for all 72 $K-$graphs
at order $N=8.$ The striking degree of similarity between
the various distributions makes plausible the existence
of a single representative function $\tilde{p}$ such
that $\int d\tilde{U} \tilde{p}/\tilde{U}^2 \simeq
\int d\tilde{U} \tilde{p_i}/\tilde{U}^2$ is
independent of the graph under consideration.
Limited information on $\tilde{p}$ can be gleaned by looking
at the moments of $\tilde{p}_i.$  For relatively small
values of $N$ ($N=6,8$) we have calculated connected moments
of $\tilde{p}_i$ and find that these
fall very rapidly compared to disconnected moments; this is
evidence that $\tilde{p}$ is quite Gaussian
above some value which near is
or below $\tilde{U} =1.$ Not much can be learned about the
small-$\tilde{U}$ behavior of $\tilde{p}$ from direct numerical
calculation; all that we know for sure is that
$\tilde{p} \simeq \tilde{U}^2$ when $\tilde{U}$ is small, since by
arguments like those given in connection with Eq.~\ref{eq:scale},
the moments $\langle \tilde{U}^k \rangle$ of $\tilde{p}$
exist for $k > -3$ and diverge at $k=-3.$ Perhaps
$\tilde{p}$ is $\tilde{U}^2$ times a Gaussian. For comparison,
the solid line of Fig.~\ref{fi:72} shows the rescaled
probability density corresponding to
the completely symmetric $U-$function.

Summing over all $K-$graphs, we can use the hypothesized density
$\tilde{p}$ to write
\begin{eqnarray}
\sum I_i & \simeq &
\displaystyle{ ( 3 N - 4 )!^2 \over ( 2 N - 3 )! } \,\,
\left( \int d\tilde U \displaystyle{ \tilde{p} \over \tilde{U}^2} \right) \,\,
\sum_i  \displaystyle{ 1 \over C_i^2 }  \nonumber \\
  & = &
\displaystyle{ ( 3 N - 4 )!^2 \over ( 2 N - 3 )! } \,\,
\displaystyle{ N! \over 2^N } \,\,
\left( \int d\tilde U \displaystyle{ \tilde{p} \over \tilde{U}^2} \right)
\left\langle \displaystyle{ 1 \over C^2 } \right\rangle .
\end{eqnarray}
In other words, the contributions of $K-$graphs to
$Im~T_{2 \rightarrow 2}$ appears to be controlled by
by two distributions: the distribution of complexities
and a characteristic probability density $\tilde{p}.$

\section{Factoring $K-$Graphs From Chains}
\label{se:proof}
\hspace*{\parindent}
Here we demonstrate the steps leading to the
factorization theorem illustrated in Fig.~\ref{fi:factor}.
For the purpose of this appendix let us assume
that the diagram of Fig.~\ref{fi:factor}a gives an
$N^{\rm th}$ order contribution to $Im~T_{2 \rightarrow 2}$
which we denote by
\begin{eqnarray}
\label{eq:recallimt}
\left( Im~T_{2 \rightarrow 2} \right)_{\rm{Fig.~\ref{fi:factor}a}}
& = & \displaystyle{g^2 \over 16 \pi }
       \left( \displaystyle{ g \over 16 \pi^2 } \right)^{N-2}
       \displaystyle{ 1 \over S }
\int^1_0 \displaystyle{ dx_1 \, dx_2 \, \prod dy_i \, \prod dz_i
\over U^2 }
\nonumber \\
& & \times \,
\delta\left( 1 - x_1 - x_2 - \sum y_i - \sum z_i \right)
\, \left[ \Theta\!\left(
\displaystyle{ \phi \over U} - m^2 \right) +
\Theta\!\left( -\displaystyle{\phi \over U} - m^2 \right)
\right] . \nonumber \\& &
\end{eqnarray}
We let $y_i$ label the internal lines of the $K-$graph,
$x_1$ and $x_2$ label the vertical lines
connecting the $K-$graph to the rest of the graph,
and the $z_i$ label the internal lines of the circular blob.
The second $\Theta-$function in Eq.~\ref{eq:recallimt} assumes
that Fig.~\ref{fi:factor}a includes
the appropriate $u-$channel exchange graph.

Let $U'$ and $\phi'$ denote the functions of $z_i$ which
characterize the circular blob in Fig.~\ref{fi:factor}a
 and let $U_K$ and
$\phi_K$ be the corresponding functions of $y_i$ for
the $K-$graph. If $U$ and $\phi$ describe the whole graph,
one can show that
\begin{equation}
\label{eq:phifac}
\phi = s \,\,
\displaystyle{  \phi_{K}         \over s }    \,\,
\displaystyle{  \phi' \over s }  \,\, ,
\end{equation}
and
\begin{equation}
\label{eq:ufac}
U = U_K \,  ( x_1 + x_2 ) \, U' + R \,\, .
\end{equation}
The remainder $R \ge 0 $ is a polynomial with the property that
if one introduces scaling variables for all the loops in
the original graph then $R$ is of higher order
in the scaling variables (see Eqs.~\ref{eq:scale0},\ref{eq:scale}).
Our strategy is to use Eqs.~\ref{eq:phifac}-\ref{eq:ufac}
to decompose $U$, the $\Theta-$function and the $\delta-$function
in Eq.~\ref{eq:recallimt} in order to demonstrate
factorization.

Since the dominant contributions to $Im~T_{2 \rightarrow 2}$
come from the region where $U$ vanishes, it is a good first approximation
to drop $R$ in Eq.~\ref{eq:ufac} and use $U = U_K \, ( x_1 + x_2 )
\, U'$ in the $1/U^2$ factor of Eq.~\ref{eq:recallimt}. The sum of
the $\Theta-$functions may be rewritten as
\begin{eqnarray}
\Theta\!\left( \displaystyle{ \phi \over U} - m^2 \right) +
\Theta\!\left(- \displaystyle{ \phi \over U} - m^2 \right)
 & = &  \Theta\! \left(\displaystyle{ \phi' / s \over ( x_1 + x_2 ) U'}
 - \displaystyle{m^2 \over s }
\left| \displaystyle{ s \over \phi_K } \right|
\left( U_K + \displaystyle{ R \over (x_1+x_2) U'} \right) \right)
\nonumber \\  & & \nonumber \\ & & \quad \quad + \quad
\label{eq:thetadecomp}
\left( \phi' \rightarrow -\phi' \right)   \, \, .
\end{eqnarray}
In the leading-log approximation the factor multiplying $m^2/s$
in Eq.~\ref{eq:thetadecomp}, namely,
\begin{equation}
\label{eq:factor}
\left| \displaystyle{ s \over \phi_K } \right|
\left( U_K + \displaystyle{ R \over (x_1+x_2) U'} \right)
\end{equation}
may be absorbed into the definition of $m^2$ since
any finite positive multiple of $m^2$ is equivalent to $m^2$
for the purpose of regulating potential logarithmic divergences
of the circular blob. To see that the inherent properties
of $K-$graphs ensure that the factor in Eq.~\ref{eq:factor}
is well-behaved, imagine rescaling the Feynman parameters
of an $l-$line loop of the $K-$graph by introducing into
Eq.~\ref{eq:recallimt} the identity $1 = \int_0^1 d\lambda \, \
\delta( \sum_{i=1}^l y_i )$ and letting $y_i = \lambda \tilde{y}_i.$
At worst, the factor of Eq.~\ref{eq:factor} scales as
$\sim 1/\lambda$ for small $\lambda,$ but this region is not
weighted heavily since the overall $\lambda-$integral varies as
$\int_0^1 \lambda^{l-3} d\lambda $ since $K-$graphs have no two-line
loops. Consequently, absorbing the $K-$graph factor of
Eq.~\ref{eq:factor} into the definition of $m^2$ amounts to writing
\begin{eqnarray}
\Theta\!\left( \displaystyle{ \phi \over U} - m^2 \right) +
\Theta\!\left(- \displaystyle{ \phi \over U} - m^2 \right)
 & \simeq &
\left[
 \Theta\! \left(\displaystyle{ \phi' / s \over ( x_1 + x_2 ) U'}
 - \displaystyle{m^2 \over s }
 \right)
+
 \Theta\! \left(- \displaystyle{ \phi' / s \over ( x_1 + x_2 ) U'}
 - \displaystyle{m^2 \over s }
 \right)
\right] \nonumber \\
 & & \nonumber \\
& & \qquad \times \,
\left[
\Theta \left( \displaystyle{   \phi_K \over U_K }\right) +
\Theta \left( - \displaystyle{ \phi_K \over U_K }\right) \right]
\end{eqnarray}
where, for future reference, we suggestively include
the second factor in square brackets which trivially sums
to unity.

Turning finally to the $\delta-$function of Eq.~\ref{eq:recallimt}
we invoke the decomposition
\begin{equation}
\label{eq:deltadef}
\delta \left( 1 - x_1 - x_2 - \sum y_i - \sum z_i \right)
 = \int_0^1 \, \, d\lambda\,\,  \delta\!\left( \lambda - \sum y_i \right)
\delta\!\left( 1 - x_1 - x_2 - \lambda - \sum z_i \right) \, . \nonumber \\
\end{equation}
Substituting Eqs.~\ref{eq:phifac}--\ref{eq:deltadef} into
Eq.~\ref{eq:recallimt} and changing variables to $y_i =
\lambda \tilde{y}_i$ (i.e., simulataneously scaling all
Feynman parameters of the $K-$graph) gives
\begin{eqnarray}
\label{eq:together} \left( Im~T_{2 \rightarrow 2}
\right)_{\rm Fig.~\ref{fi:factor}a} & = & \displaystyle{g^2 \over 16 \pi }
\left( \displaystyle{ g \over 16 \pi^2 } \right)^{N-2}
\displaystyle{ 1 \over S } \int^1_0
\displaystyle{ \prod d\tilde{y}_i \,
\delta\left( 1 - \sum \tilde{y}_i \right)
\over U^2_K(\tilde{y}) } \nonumber \\
& & \nonumber \\ & &  \qquad \qquad \qquad \times \quad \int^1_0
\displaystyle{ d\lambda \, dx_1 \, dx_2 \prod dz_i \,
\delta \left( 1 - x_1 - x_2 - \lambda - \sum z_i \right)
\over \lambda \, \left( (x_1 + x_2) U' \right)^2 }
\,\, \nonumber \\ & & \nonumber \\  & & \qquad \qquad \qquad \qquad \times \,
\left[ \Theta\! \left( \displaystyle{ \phi'/s \over ( x_1 + x_2 ) U'}
- \displaystyle{m^2 \over s } \right) \quad + \quad \phi'
\rightarrow -\phi'\right] \nonumber \\ & &
\end{eqnarray}
where the integral over $\tilde{y}_i$ defines an overall multiplicative
factor because
\begin{equation}
\displaystyle{\prod dy_i \over  U^2_{K}(y_i)} =
\displaystyle{ \prod d\tilde{y}_i \over  U^2_{K}(\tilde{y}_i) } \, ,
\end{equation}
is independent of $\lambda.$ The integrals over $\lambda,
x_1,x_2$ and $z_i$ in Eq.~\ref{eq:together} may be put
in a familiar form by transforming from $\lambda$ to
the variables $w_1$ and $w_2$ through
\begin{equation}
\label{eq:wdef}
\int_0^1
\displaystyle{ d\lambda \over \lambda  }
\,\, \delta\!\left( 1 - x_1 - x_2 - \lambda - \sum z_i \right)
= \int_0^1
\frac{ dw_1 dw_2 }{ ( w_1 + w_2 )^2  }
\, \delta \left( 1 - x_1 - x_2 - w_1 - w_2 - \sum z_i \right) .
\end{equation}
This transformation is most easily verified by inserting
$1=\int_0^1d\lambda\, \delta(\lambda- w_1 -w_2)$ on the
right-hand side and then performing a change of variables
by defining $w_i = \lambda \tilde{w}_i.$

Before putting together all the above expressions,
consider the graph of Fig.~\ref{fi:factor}b (disregarding
the factor $\hat{K}$ which we will discuss below) obtained
by replacing the $K-$graph of Fig.~\ref{fi:factor}a with a two-line
loop labelled by the Feynman parameters $w_1$ and $w_2.$
If we let $\phi''$ and $U''$ denote the functions characterizing
Fig.~\ref{fi:factor}b then
\begin{equation}
\label{eq:newphi}
\phi'' = w_1 w_2 \,\, \displaystyle{ \phi' \over s } \,\, ,
\qquad \qquad
U'' = (w_1 + w_2) (x_1 + x_2 ) U' + R'' \, ,
\end{equation}
where the remainder  $R''$ is analogous to that in Eq.~\ref{eq:ufac}.
To leading-log accuracy the $\Theta-$functions
in Eq.~\ref{eq:together} can be re-expressed in terms of
$\phi''$ and $U''$ because
\begin{equation}
\label{eq:newtheta1}
\Theta\! \left( \displaystyle{ \phi'/s
\over ( x_1 + x_2 ) U' }
 - \displaystyle{m^2 \over s } \right)
 \simeq
\Theta\! \left( \displaystyle{ \phi'/s
\over ( x_1 + x_2 ) U'  }
 - \displaystyle{m^2 \over s } \displaystyle{ (w_1 + w_2 ) \over w_1 w_2 }
\right)   =
\Theta\! \left( \displaystyle{ \phi'' \over U'' }
 - m^2
\right)
\end{equation}

Substituting Eqs.~\ref{eq:wdef}-\ref{eq:newtheta1} into
Eq.~\ref{eq:together} we arrive at the final result
\begin{equation}
  \left( Im~T_{2 \rightarrow 2 } \right)_{\rm{Fig.~\ref{fi:factor}a}}
= \hat{K} \, \times \,
\left( Im~T_{2 \rightarrow 2 } \right)_{\rm{Fig.~\ref{fi:factor}b}}
\end{equation}
where for a $N_K^{\rm th}$  order $K-$graph,
\begin{equation}
\hat{K} \equiv
\displaystyle{ \left( Im~T_{2 \rightarrow 2 } \right)_K \over
               \left( Im~T_{2 \rightarrow 2 } \right)_{N=2} }
=
\displaystyle{
\displaystyle{g^2 \over 16 \pi }
\left( \displaystyle{ g \over 16 \pi^2 } \right)^{N_K -2}
\int^1_0 \displaystyle{ [d\tilde{y}] \over U^2_K }
\over
\displaystyle{g^2 \over 16 \pi } \,\, \frac{1}{2}
}
= 2
\left( \displaystyle{  g \over 16 \pi^2 } \right)^{N_K -2}
\int^1_0 \displaystyle{ [d\tilde{y}] \over U^2_K }  \,\, .
\end{equation}
The denominator $\left( Im~T_{2 \rightarrow 2 } \right)_{N=2}$
in the definition of $\hat{K}$ is simply the imaginary part
of a single two-line loop.

The contribution of the graph of Fig.~\ref{fi:factor}b is
\begin{eqnarray}
 \left( Im~T_{2 \rightarrow 2 } \right)_{\rm{Fig.~\ref{fi:factor}b}}
& = & \displaystyle{g^2 \over 16 \pi }
       \left( \displaystyle{ g \over 16 \pi^2 } \right)^{N-N_K-2}
       \displaystyle{ 1 \over 2 S }
\int^1_0
\displaystyle{ dw_1 \, dw_2 \, dx_1 \, dx_2 \prod dz_i
\over  U''^2 } \nonumber \\
& & \nonumber \\
& &  \times \,
\delta \!\left( 1 - x_1 - x_2 - w_1 - w_2 - \sum z_i \right)
\left[
\Theta\! \left( \displaystyle{ \phi'' \over U'' }
 - m^2
\right) \, + \, \left(
\phi'' \rightarrow -\phi'' \right)
\right] \, ,
\nonumber \\
 & &
\end{eqnarray}
where the factor of $2$ accompanying the symmetry factor $S$
(which describes Fig.~\ref{fi:factor}a) accounts for
extra two-line loop. To keep the discussion simple, we
have implicitly assumed throughout the discussion
that the isolated $K-$graph has a symmetry factor of unity.

	One might wonder how general the analysis of
Fig.~\ref{fi:factor} can be. It turns out that imaginary
part of every graph we consider can be written, even with
two legs off-shell, as a  linear superposition of the
the imaginary parts of  one-loop graphs. Consider the
graph of Fig.~\ref{fi:dgs}, where it is understood
that the internal lines have mass $M$
(not to be confused with any physical mass $m$). Note particularly
that the momentum $p'$
has been replaced by $\beta p',$ where $\beta,$ it turns out, runs
from $-1$ to 1. Actually, only $\beta > 0$ contributes to the imaginary
part,
as we will see. The graph of Fig.~\ref{fi:dgs} has the value
(aside from an additive
cut-off dependent contribution to its real part)
\begin{equation}
\int_0^1 dx \, \ln \left[ M^2 - x ( 1- x ) ( - k + \beta p')^2 \right] .
\end{equation}
Multiply this by a function $F(M^2,\beta),$ whose significance we
reveal below, take the imaginary part, and
integrate over $M^2 > 0$, $|\beta| \le 1$;
the result is
\begin{equation}
\label{eq:decomp}
2 \pi \int^\infty_0 dM^2
\int^1_{-1} d\beta \, F(M^2,\beta)\int_0^{1/2} dx\,
\Theta\left[ ( - k + \beta p')^2 - \frac{ M^2 }{ x(1-x) } \right]
\end{equation}
In writing eq.~\ref{eq:decomp} we have made use of the symmetry
in $x $ and  $1-x.$ Now change variables to
\begin{equation}
\rho^2 = \frac{M^2}{x(1-x)},\qquad \qquad dx = \frac{M^2}{\rho^4} d\rho^2
\left( \rho^2 - 4 m^2 \right)^{-1/2}
\end{equation}
Then Eq.~\ref{eq:decomp} becomes
\begin{equation}
\label{eq:hdef}
 \int^\infty_0 d\rho^2 \int^1_{-1} d\beta \, h(\rho^2,\beta)
\, \Theta\left[ ( - k + \beta p')^2 - \rho^2  \right]
\end{equation}
with
\begin{equation}
h(\rho^2,\beta) = \frac{2 \pi}{\rho^4}
\int_0^{\rho^2 / 4} dM^2 \, M^2 F(M^2, \eta)
\left( \rho^2 - 4 M^2 \right)^{-1/2}.
\end{equation}
This relation between $h$ and $F$ is Abel's integral equation which
can be explicitly solved.

	The significance of Eq.~\ref{eq:hdef}  is that it is a general
representation\cite{dese,nak} of the
imaginary part of the Feynman graphs we
consider; we call it (following ancient usage\cite{corn3}) the
DGS representation. It is a form of the Jost-Lehmann-Dyson representation
which can be derived either by appealing to causality and spectrum
conditions\cite{dese}, or by direct investigation of Feynman-parameter
representations of graphs\cite{nak}. The DGS representation
reduces the analysis of general graphs to a linear superposition
of Fig.~\ref{fi:dgs} (with $p' \rightarrow \beta p'$).
We note that the physical region of $Im~T(-k,\beta p')$
is restricted by the requirement $-k_0 + \beta p_0' \ge 0,$
which turns out to mean $\beta \ge 0.$
This is just telling us that $Im~T$ receives contributions
from parts involving $\ln(-s/m^2),$ not $\ln(-u/m^2).$

\section{Bounds From Beyond $K-$graphs}
\label{se:beyondk}

\hspace*{\parindent}
Our estimates of the contributions of
pure $K-$graphs (and their exchange graphs) to
$Im~T_{2 \rightarrow 2}$ involved the complexity
density function multiplied by the total number
of $K-$graphs. In this section we extend our estimates
by considering an even larger class of graphs without
logarithms which encompasses both $K-$graphs, their
exchange graphs, and $s-$channel graphs.

Of use to us is a combinatorial result of Bender
and Canfield\cite{bencan} who obtained an expression
for the asymptotic number of labelled graphs with a
specified number of lines attached to each vertex.
In particular we are interested in the number
of $N^{\rm th}$ order $\phi^4$ graphs with $4$ external
legs such that the graphs have neither tadpoles nor
multiple lines between vertices ---
these criteria ensure the absence of logarithms
in the corresponding contributions to
$Im~T_{2 \rightarrow 2}.$ The number of labelled graphs
with the desired properties is asymptotically equal to
\begin{equation}
\label{eq:diag}
\displaystyle{ (4 N + 4 )! \over ( 2 N + 2 )! } \,\,
\displaystyle{ 1\over 2^{2 N + 2} } \,\,
\displaystyle{ 1 \over (4!)^N }\,\,
\exp\left[
- \displaystyle{ 3 N \over 2 ( N + 1 ) } -
\left( \displaystyle{ 3 N \over 2 ( N + 1 ) } \right)^2
\right]
\doteq
\left( \displaystyle{ 2 \over 3 } \right)^N (N!)^2 \, \, .
\end{equation}
Converting from the language of labelled graphs to the
the usual implementation of the Feynman rules which
employs unlabeled graphs amount to dividing the
Eq.~\ref{eq:diag} by $N!$ so that the the effective number
of diagrams without logarithms is $\doteq (2/3)^N N!.$
For example, the corresponding bound on
$Im~T_{2 \rightarrow 2}$ analogous to
Eq.~\ref{eq:simplebounds} is
obtained simply by replacing $N!/2^N$ (the number of
$K-$graphs) with $(2/3)^N N!$ to obtain $a \ge 2/3$ in
\begin{equation}
Im~T_{2 \rightarrow 2} \doteq N!
\left(
\displaystyle{ a g \over 16 \pi^2 }
\right)^N .
\end{equation}
If, as in Appendix~B, we speculate about further improving
the bounds on $a$ by a factor of $( C_S/\langle C_i \rangle)^2
\simeq (32/27)^2$ (assuming that the average complexity
of $K-$graphs is the same as the average complexity
of all simple graphs) we obtain $a \simeq 2048/2187 \simeq 0.94.$

\section*{Figure Captions}

\begin{enumerate}

\item{\label{fi:ladder}
A multiperipheral or straight-ladder graph for $2 \rightarrow 2$
scattering in $g\phi^4$ theory. In this and all other figures
of this paper $g \phi^4$ vertices are indicated by dots.}

\item{\label{fi:tree}
Free ends of opposing multiperipheral tree graphs may be joined
in all possible ways to generate generalized $2 \rightarrow 2$ amplitudes. }

\item{\label{fi:k8}
a) An eighth-order graph with no two-line loops.
b) The $u-$channel exchange graph corresponding to a).
c) Redrawn version of b) indicates that the $u-$channel
exchange graph is not of the form of an eighth-order $K-$graph.}

\item{\label{fi:4cross}
a) A fourth-order crossed-ladder graph
b) The $u-$channel exchange graph corresponding to a).
}

\item{\label{fi:tangle}
a) The simplest example of a $K-$graph.
b) Sixth-order graph containing
a two-line loop and a $K-$graph.
c) Eighth-order graph containing
a $K-$graph crossed by two-line loops.
}

\item{\label{fi:cayley}
Two decay-like Cayley trees emerge from the point where $p$ and $p'$ meet.
Summing over all such amplitudes and squaring contributes to the
two-particle total cross section and $Im~T_{2 \rightarrow 2}.$}

\item{\label{fi:bssol}
A $2\rightarrow 2$ chain graph formed
by linking together arbitrary combinations of $K-$graphs and two-line
loops.}

\item{\label{fi:bseq}
Graphical representation of the
Bethe-Salpeter equation which sums all of
amplitudes of the form shown in Fig.~\ref{fi:bssol}.
The circular blobs include the corresponding $u-$channel
exchange graphs.}

\item{\label{fi:factor}
Illustration of the factorization theorem which permits
replacement of a $K-$graph in a chain graph with
a two-line loop and a numerical factor $\hat{K}$ given
by Eq.~\ref{eq:khatdef}. The circular blob is
assumed to include contributions from $u-$channel exchange
graphs.}

\item{\label{fi:comp}
Distribution of the complexity for all 43,200 $K-$graphs
for $N=14.$}

\item{\label{fi:asym}
Comparison of the asymptotic form of Eq.~\ref{eq:result} with numerical
integration of $\int_0^1 [dx] / U^2_S.$}

\item{\label{fi:72}
Histograms show the probability density of
$U/\langle U \rangle_i$ for all 72  eighth-order $K-$graphs.
The vertical dark bands are a consequence of the similarity
between the distributions in addition to fluctuations in the
Monte-Carlo calculation of each histogram. Solid curve
shows the corresponding distribution for the completely
symmetric function $U_S$.}

\item{\label{fi:dgs}
The off-shell two-line loop whose
weighted sum appears in the DGS representation.}

\end{enumerate}

\newpage
\oddsidemargin=1cm
\begin{center}
\begin{description}
\item[Table 1]
Average complexity of $N^{\rm th}$ order $K-$graphs. Up to
$N=12$ $C$ is calculated for all $ (N/2)! (N/2\,-1)! / 2 $
graphs but for larger $N,$ due to practical limitations of
computer time, only a reasonable number of
randomly generated graphs are averaged over.
\end{description}
\end{center}

\begin{center}
\begin{tabular}
  { | r |  l | l | r | } \hline
      &                        &                 &  \\
$N$   & ~~~~~$\langle C \rangle$   &
$\sqrt{ \langle C^2 \rangle - \langle C\rangle^2 }$ & \# graphs \\
      &                        &                 &  \\
\hline
      &                        &                 &  \\
4   &    ~~~16                  &     0                     &     1\\
6   &    1.29  $\times 10^2   $ &   1                       &     6\\
8   &    1.13 $\times 10^3    $ &  6.78 $\times 10^1$       &    72\\
10  &    1.04 $\times 10^4    $ &  8.89 $\times 10^2$       &  1440\\
12  &    9.96 $\times 10^4    $ &  1.01 $\times 10^4$       & 43200\\
20  &    1.02 $\times 10^9    $ &  1.34 $\times 10^8$       & 10000\\
30  &    1.32 $\times 10^{14} $ &  1.91 $\times 10^{13}  $  & 10000\\
40  &    1.90 $\times 10^{19} $ &  2.90 $\times 10^{18}  $  & 10000\\
50  &    2.92 $\times 10^{24} $ &  4.67 $\times 10^{23}  $  & 10000\\
60  &    4.67 $\times 10^{29} $ &  7.53 $\times 10^{28}  $  & 10000\\
70  &    7.66 $\times 10^{34} $ &  1.25 $\times 10^{34}  $  & 10000\\
100 &    3.80 $\times 10^{50} $ &  6.43 $\times 10^{49}  $  &  5000\\
200 &    1.28 $\times 10^{103}$ &  2.20 $\times 10^{102} $  &   500\\
300 &    5.67 $\times 10^{155}$ &  9.66 $\times 10^{155} $  &   500\\
400 &    2.96 $\times 10^{208}$ &  5.15 $\times 10^{207} $  &   500\\
 & & & \\
\hline
\end{tabular}
\end{center}

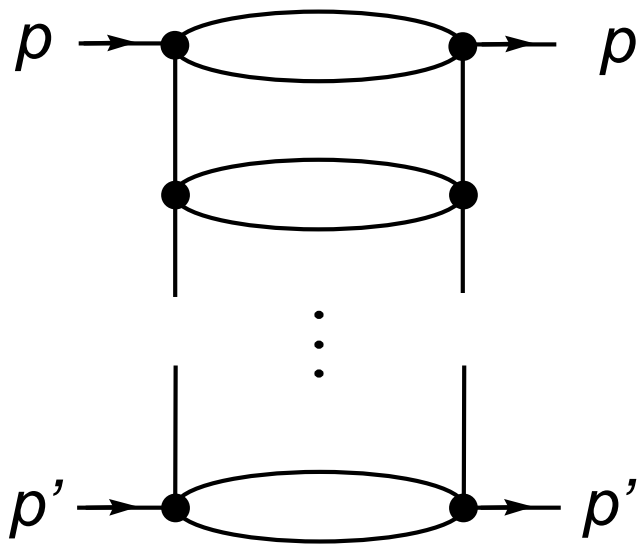
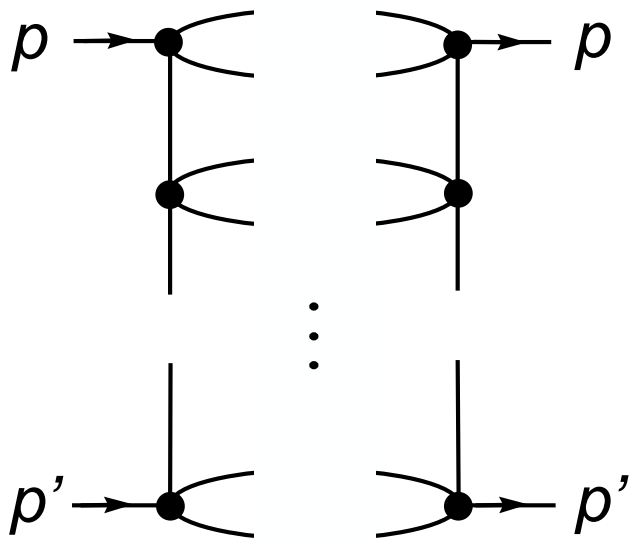
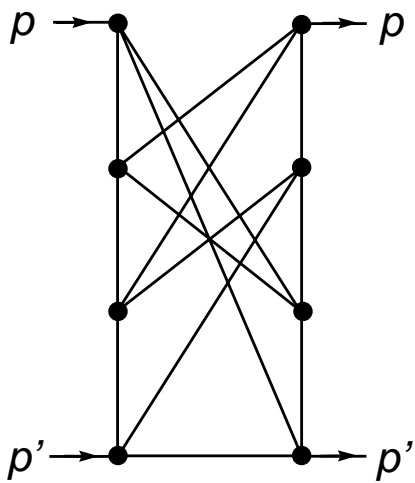
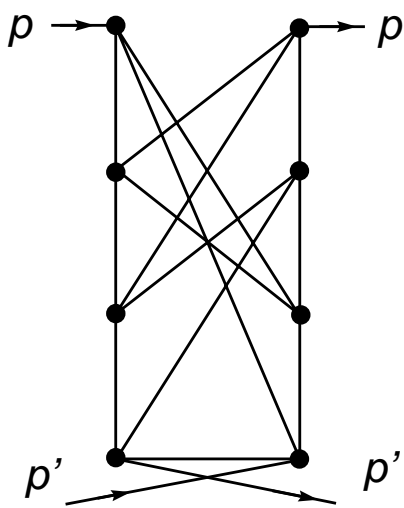
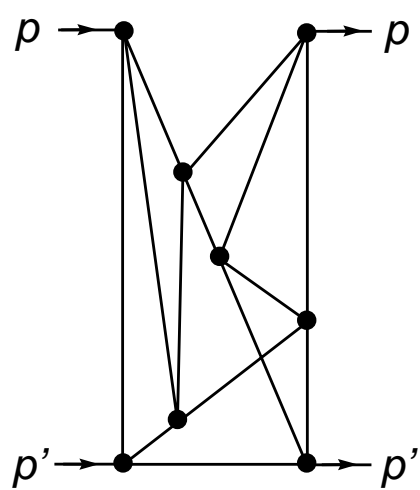
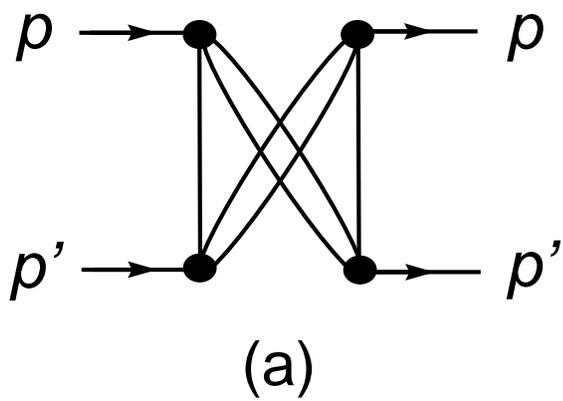
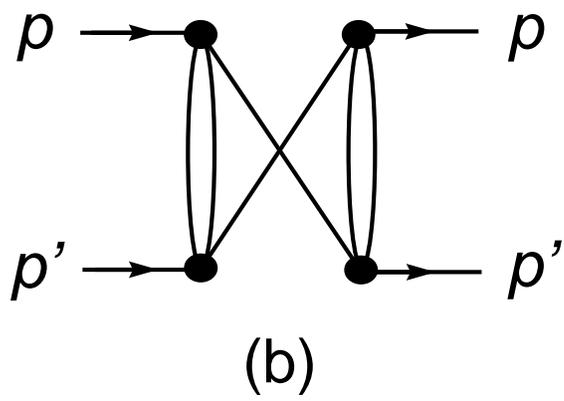
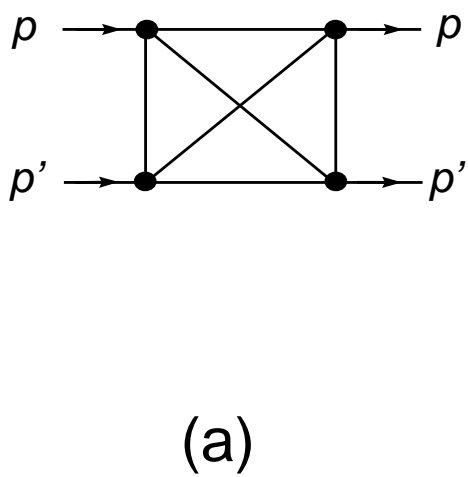
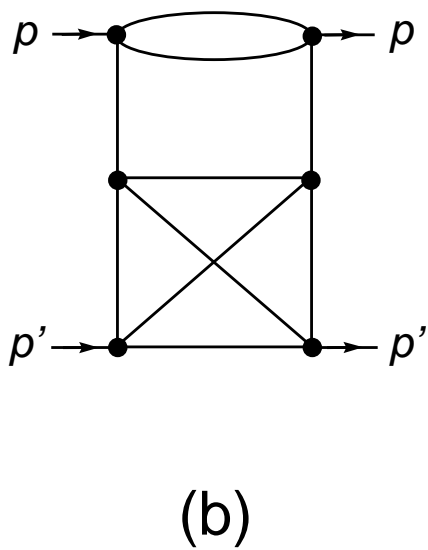
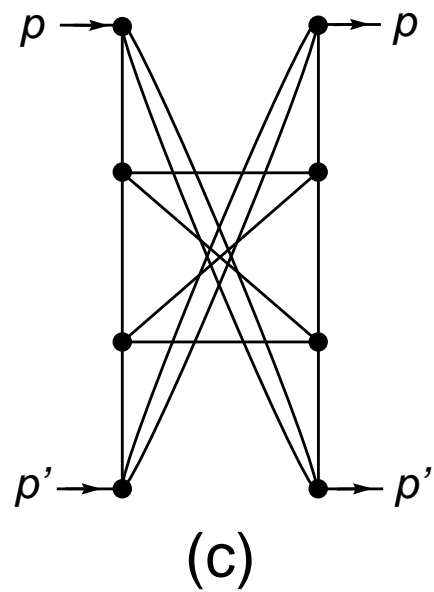
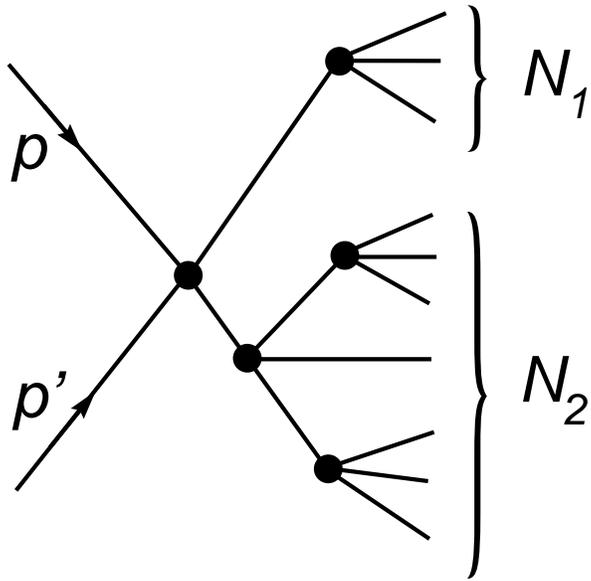
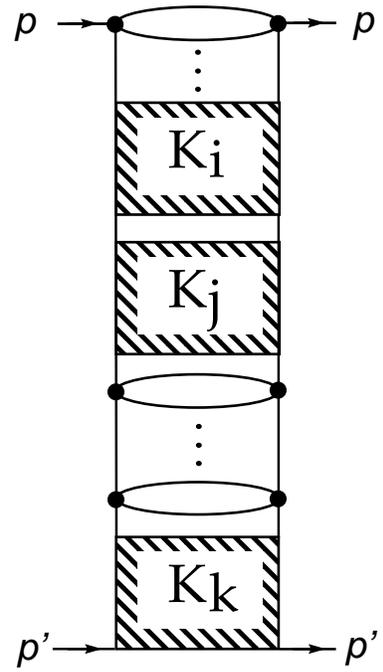
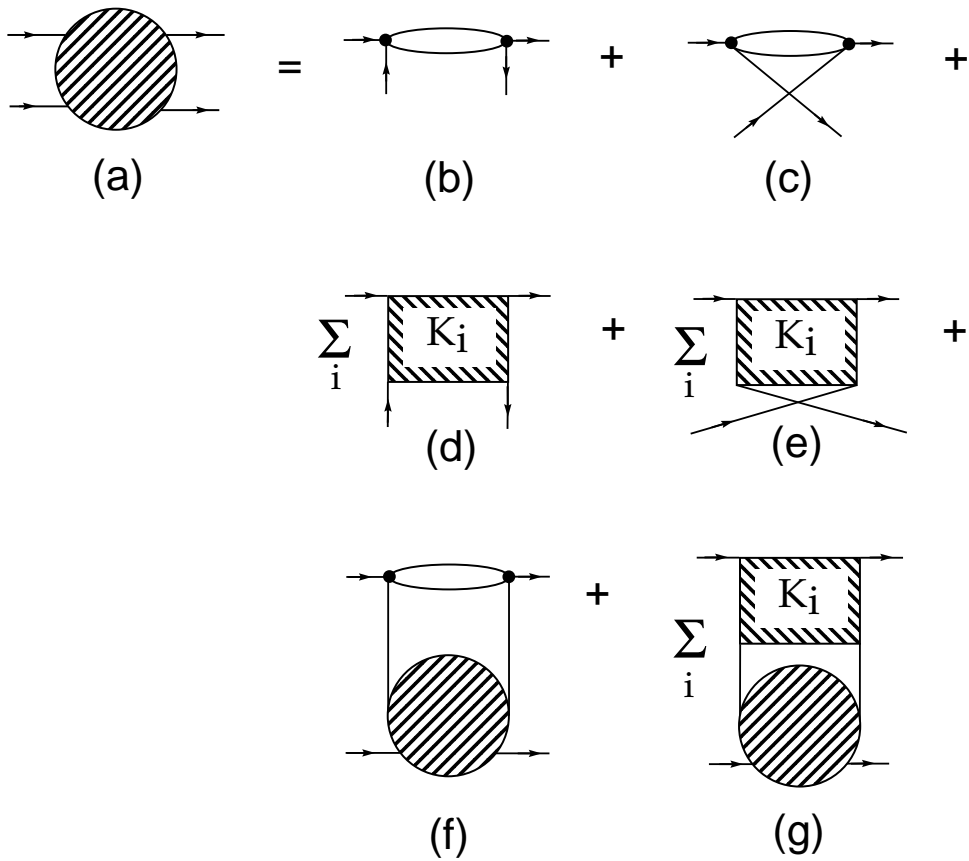
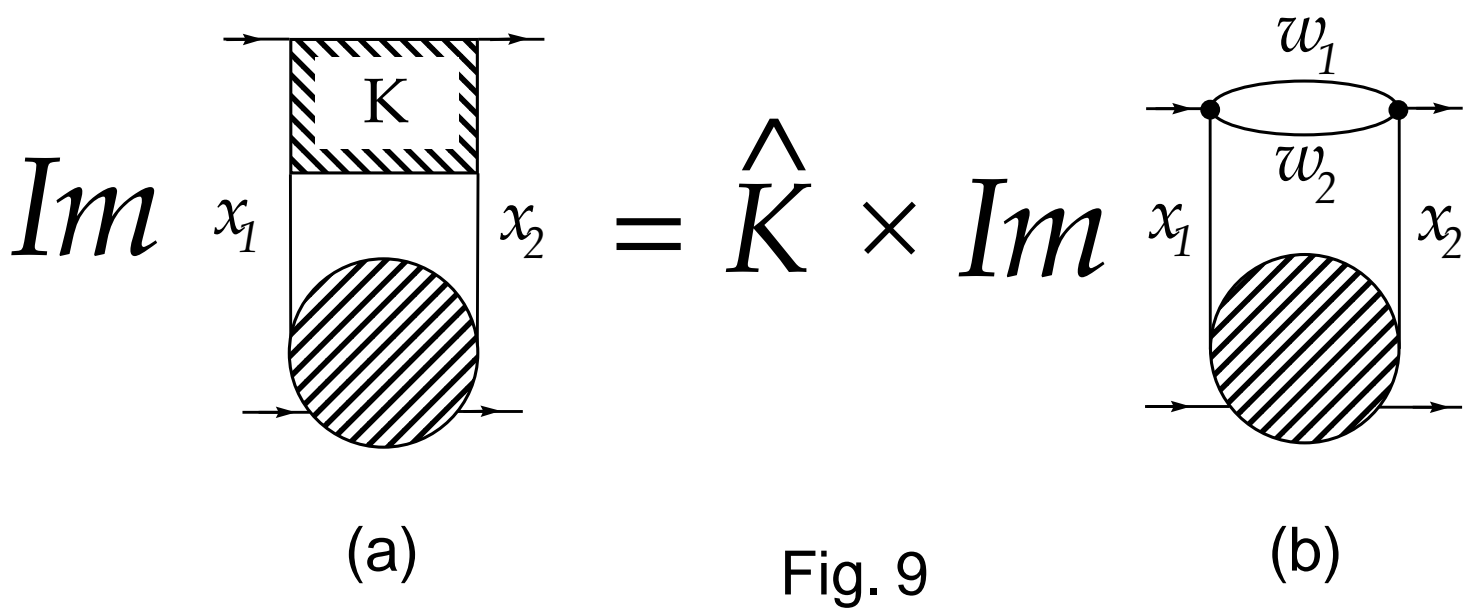
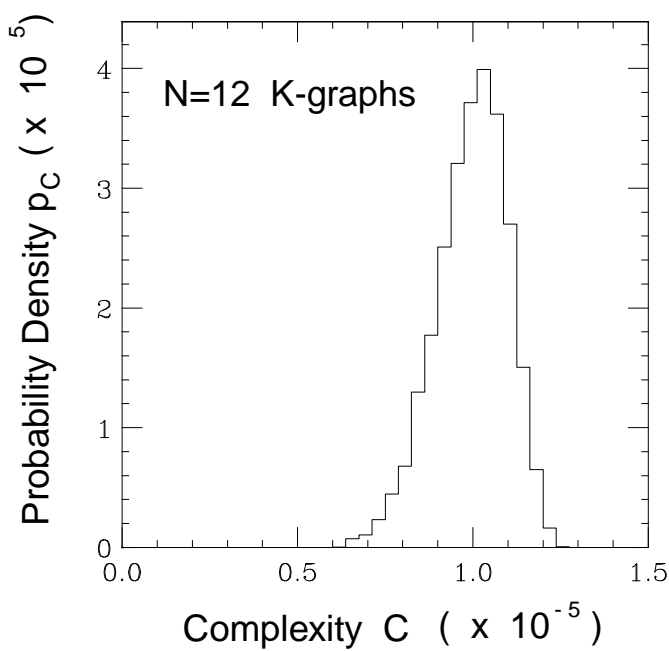
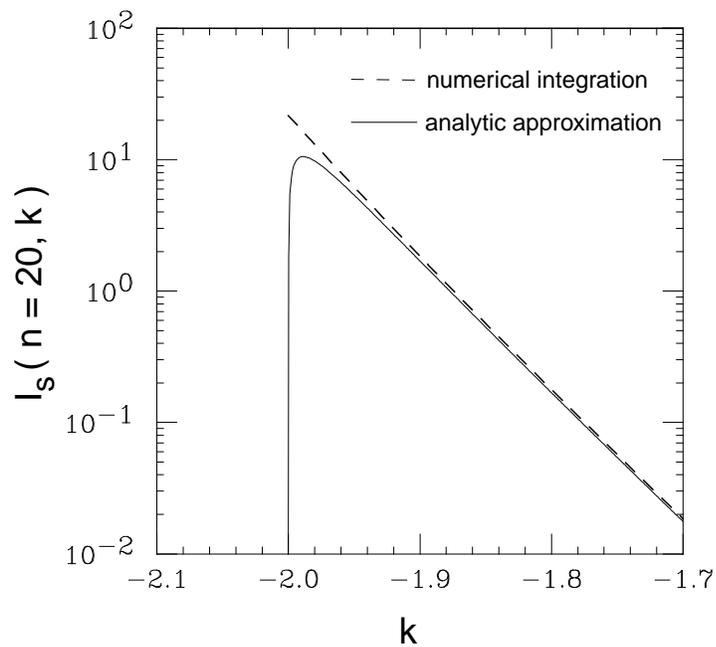
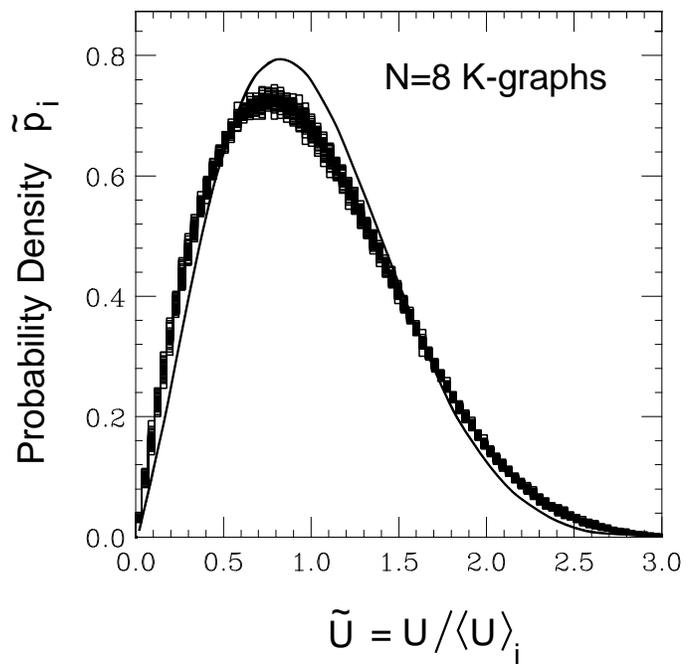
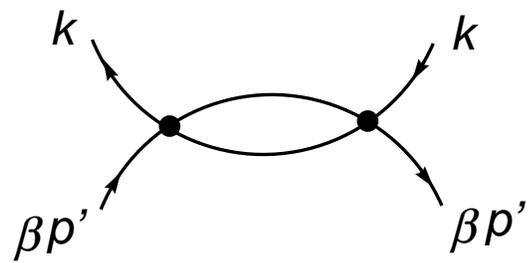
\end{document}